\documentclass[rmp,twocolumn,nofootinbib]{revtex4-1}

\usepackage{graphicx}
\usepackage{amssymb,amsmath}
\usepackage{color,soul}

\begin{document}
\title{New physics search with flavour in the LHC era}

\author{Tobias Hurth}
\email{tobias.hurth@cern.ch}
\affiliation{\mbox{PRISMA Cluster of  Excellence \& Institute for Physics (THEP),
Johannes Gutenberg University, D-55099 Mainz, Germany}} 
\author{Farvah Mahmoudi}
\email{mahmoudi@in2p3.fr}
\affiliation{\mbox{CERN Theory Division, Physics Department, CH-1211 Geneva 23, Switzerland}}
\affiliation{\mbox{Clermont Universit{\'e}, Universit\'e Blaise Pascal, CNRS/IN2P3, LPC, BP 10448, 63000 Clermont-Ferrand, France}}

\begin{abstract}
\mbox{We give a status report on quark flavour physics in view of the latest data 
from the $B$ factories and the LHC, and discuss the} \mbox{impact of the latest experimental results on new physics in the MFV framework. We also show some examples of the implica-} tions in supersymmetry.
\end{abstract}

\maketitle
 \tableofcontents

\section{SM and new physics flavour problems}
At the end of the $B$ factories at SLAC ({BaBar} experiment) (\citeauthor{Babar})
and at KEK (Belle experiment) (\citeauthor{Belle}) and of the Tevatron $B$ physics experiments
(\citeauthor{TevatronB1,TevatronB2}), 
all present measurements in flavour
physics are consistent with the simple Cabibbo-Kobayashi-Maskawa (CKM) theory of the Standard Model (SM).
The recent measurements by the high-statistics  LHCb experiment (\citeauthor{LHCb}) have not changed this feature.
Of course there have been and there are still so-called tensions, anomalies, or puzzles in the quark flavour data
at the 1-, 2-, or 3-$\sigma$ level, however, until now they all have disappeared after some time when 
more statistics had been collected.

Thus, at least at present all flavour-violating processes between quarks are well-described by a $3 \times 3$ unitarity matrix, usually referred to as the CKM
matrix \cite{Kobayashi:1973fv,Cabibbo:1963yz}, which is fully described by four real parameters, three rotation angles and one complex phase. It is this complex phase that
represents the only source of CP~violation in the SM and that 
allows for a unified description of all the CP violating
phenomena. This is an impressing success of the SM and the 
CKM theory.

\begin{figure}[!h]
\hspace*{-0.5cm}\includegraphics[width=22pc]{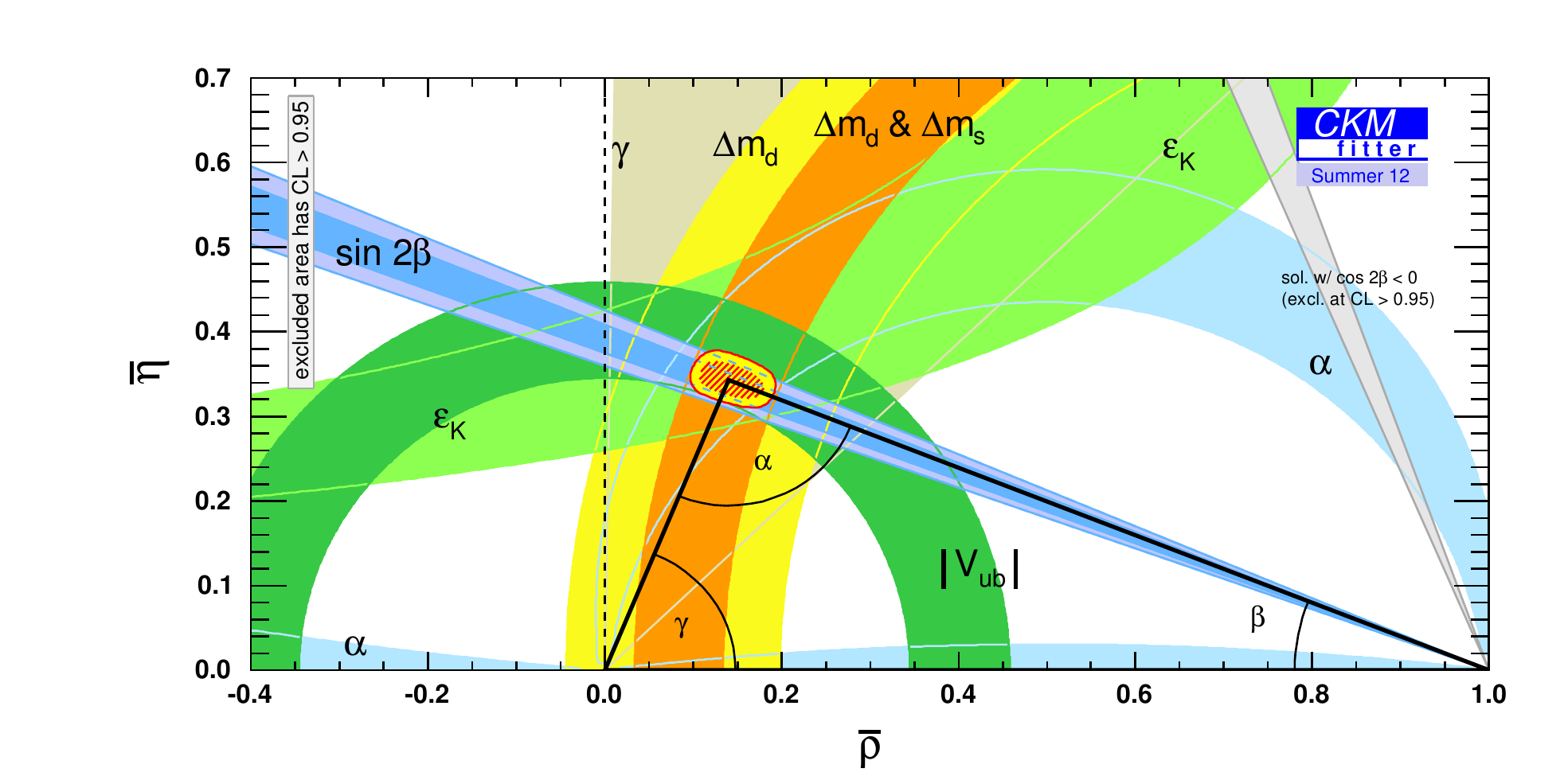}
\caption{\label{FigureCKM}Constraints in the ($\bar{\rho}$,$\bar{\eta}$)  plane. The (red) hashed region of the global combination corresponds to $68\%$ C.L.~\cite{CKMfitter}.}
\end{figure}

\begin{figure*}[!t]
\hspace*{-0.3cm}\includegraphics[width=21pc,height=10pc]{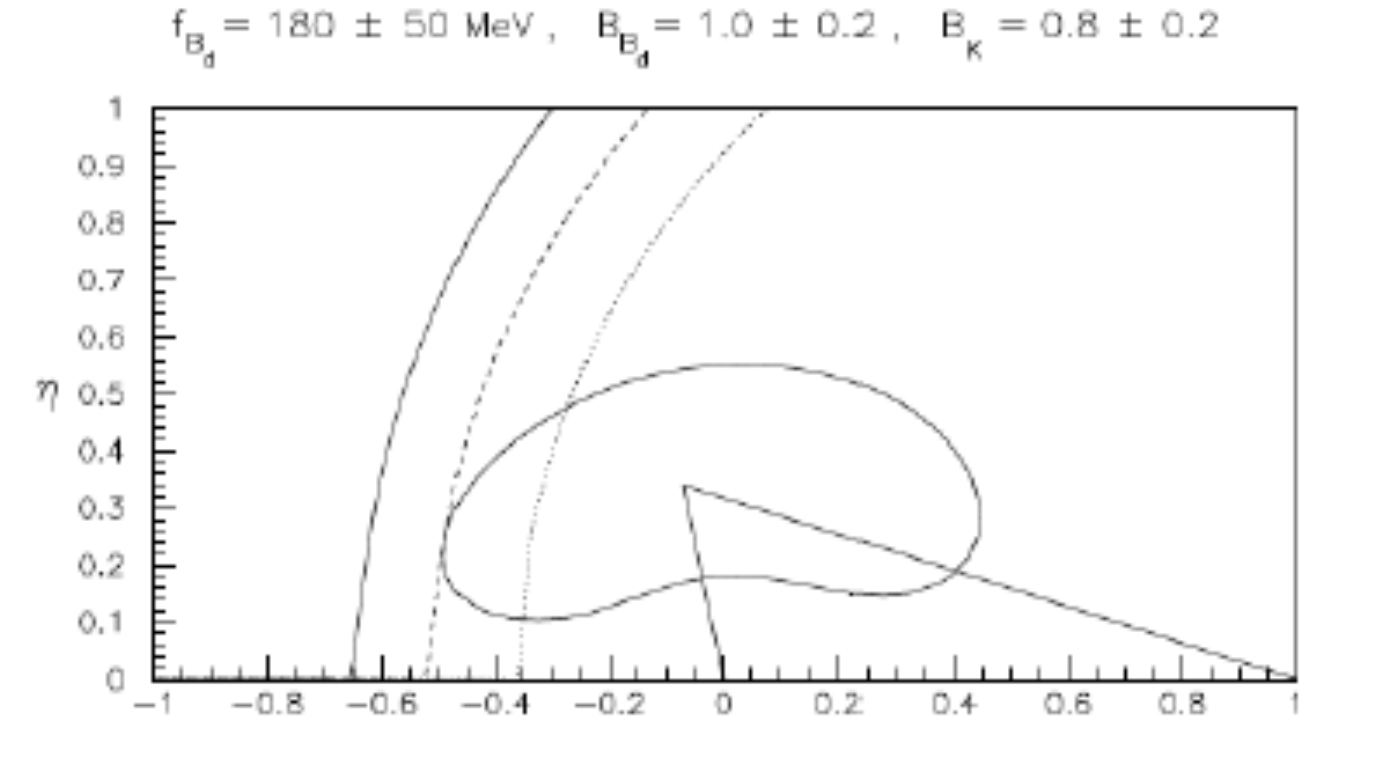}\hspace*{0.3cm}\raisebox{-0.3cm}{\includegraphics[width=21pc,height=10pc]{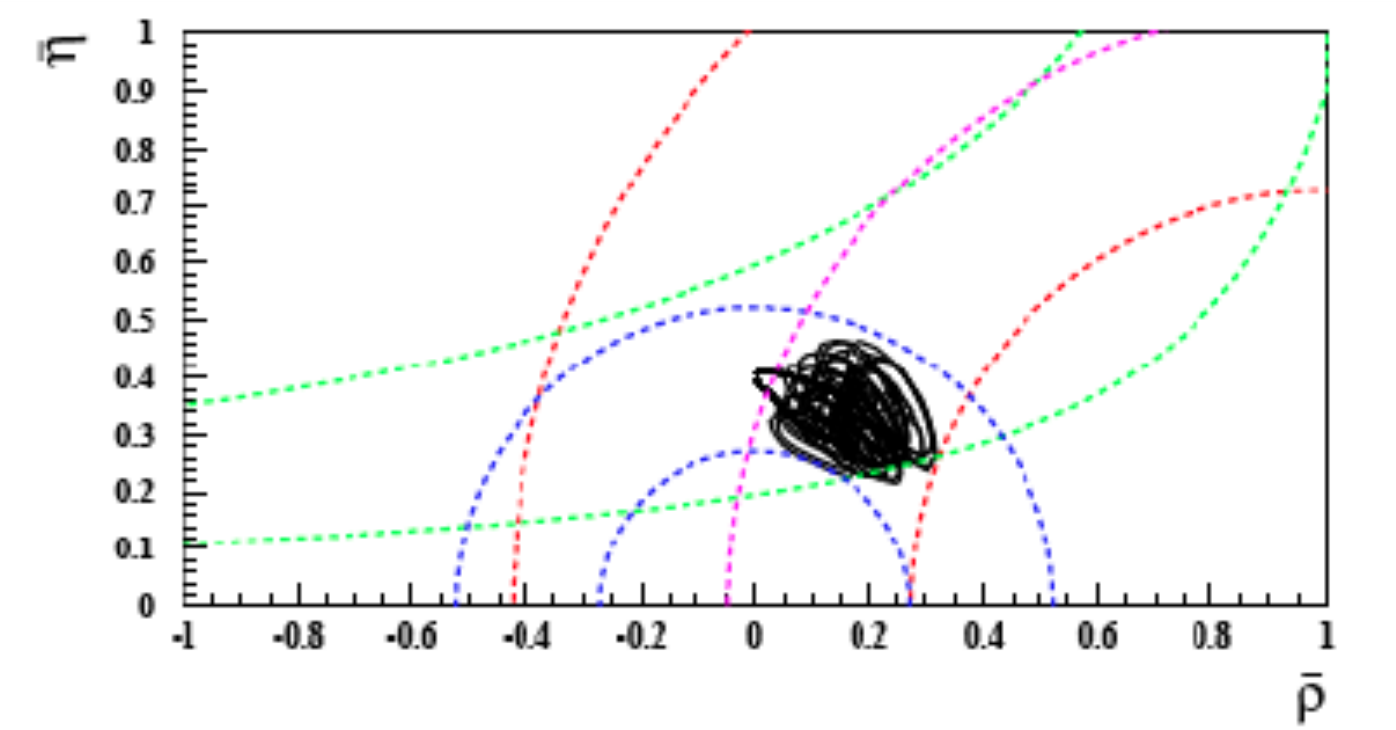}}
\caption{Historical CKM fits of (Ali and London, 1995)  (left), of Plazczynski and Schune (BABAR Physics Book, 1998) (right).} \label{FigureCKMhistoric} 
\end{figure*}

\begin{figure*}[t]
\hspace*{-0.8cm}\includegraphics[width=22pc]{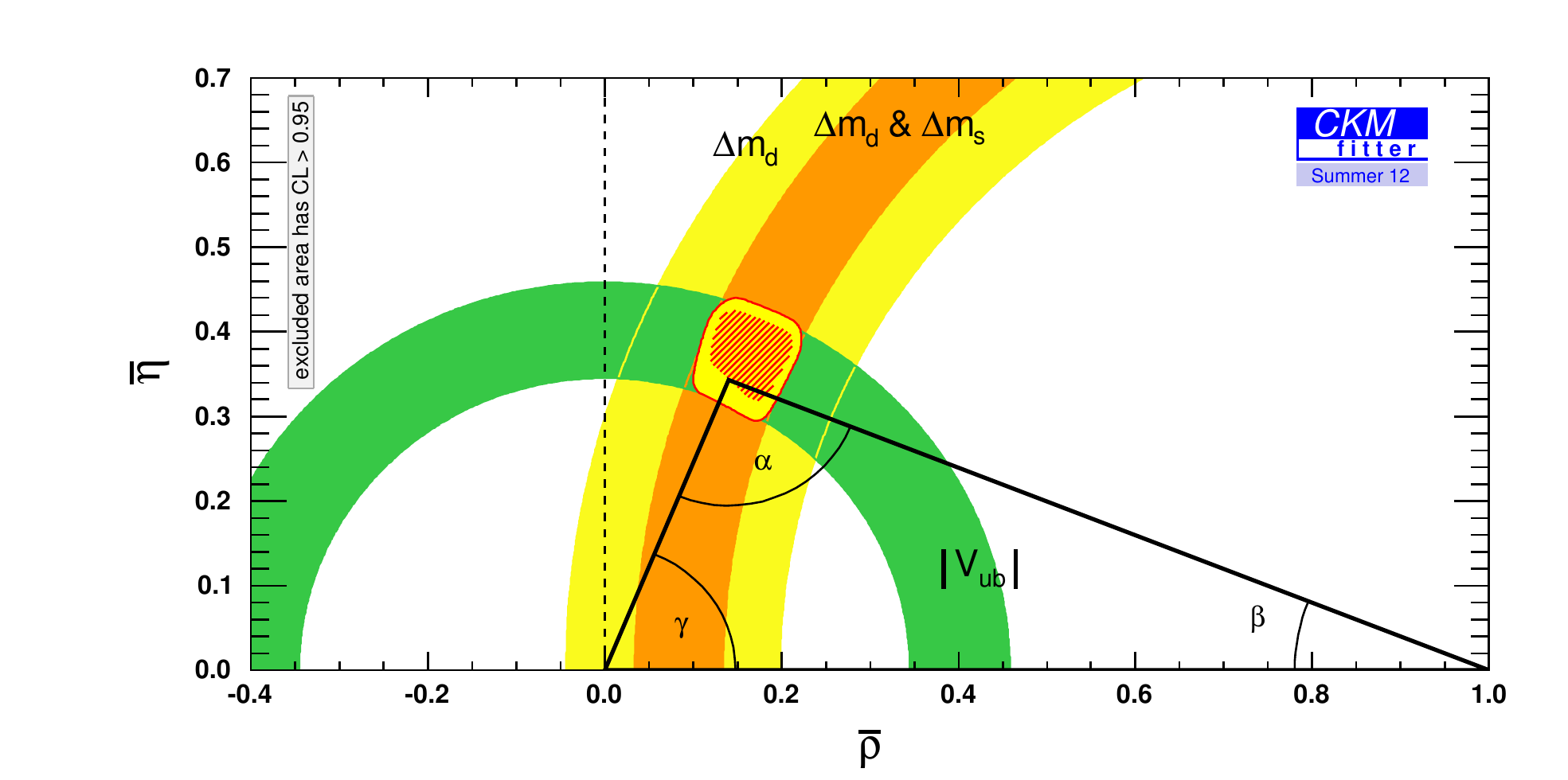}\includegraphics[width=22pc]{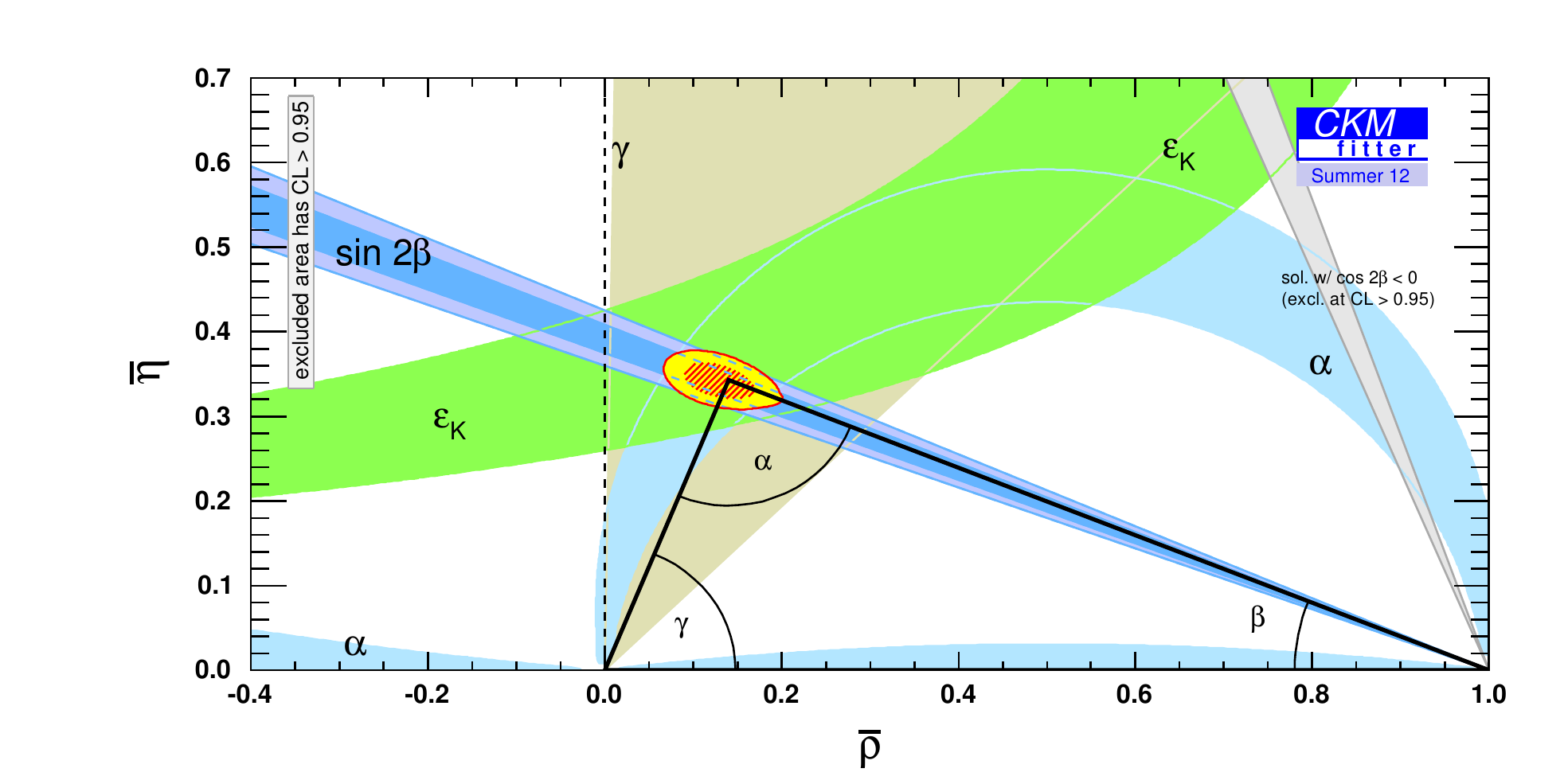}
\caption{Constraints from CP conserving (left) and CP violating (right) quantities only~\cite{CKMfitter}.}\label{FigureCKMCP2} 
\hspace*{-0.8cm}\includegraphics[width=22pc]{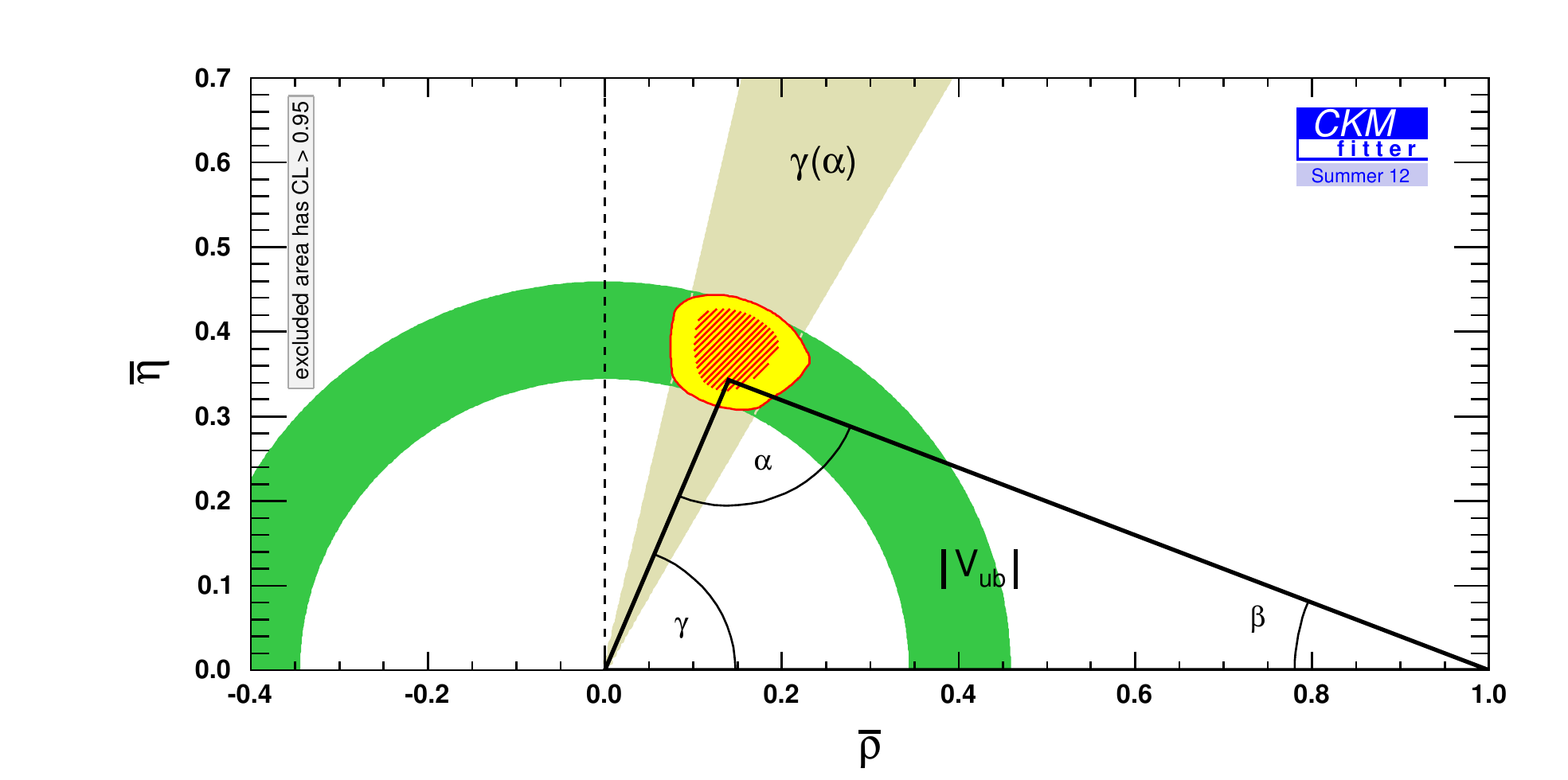}\includegraphics[width=22pc]{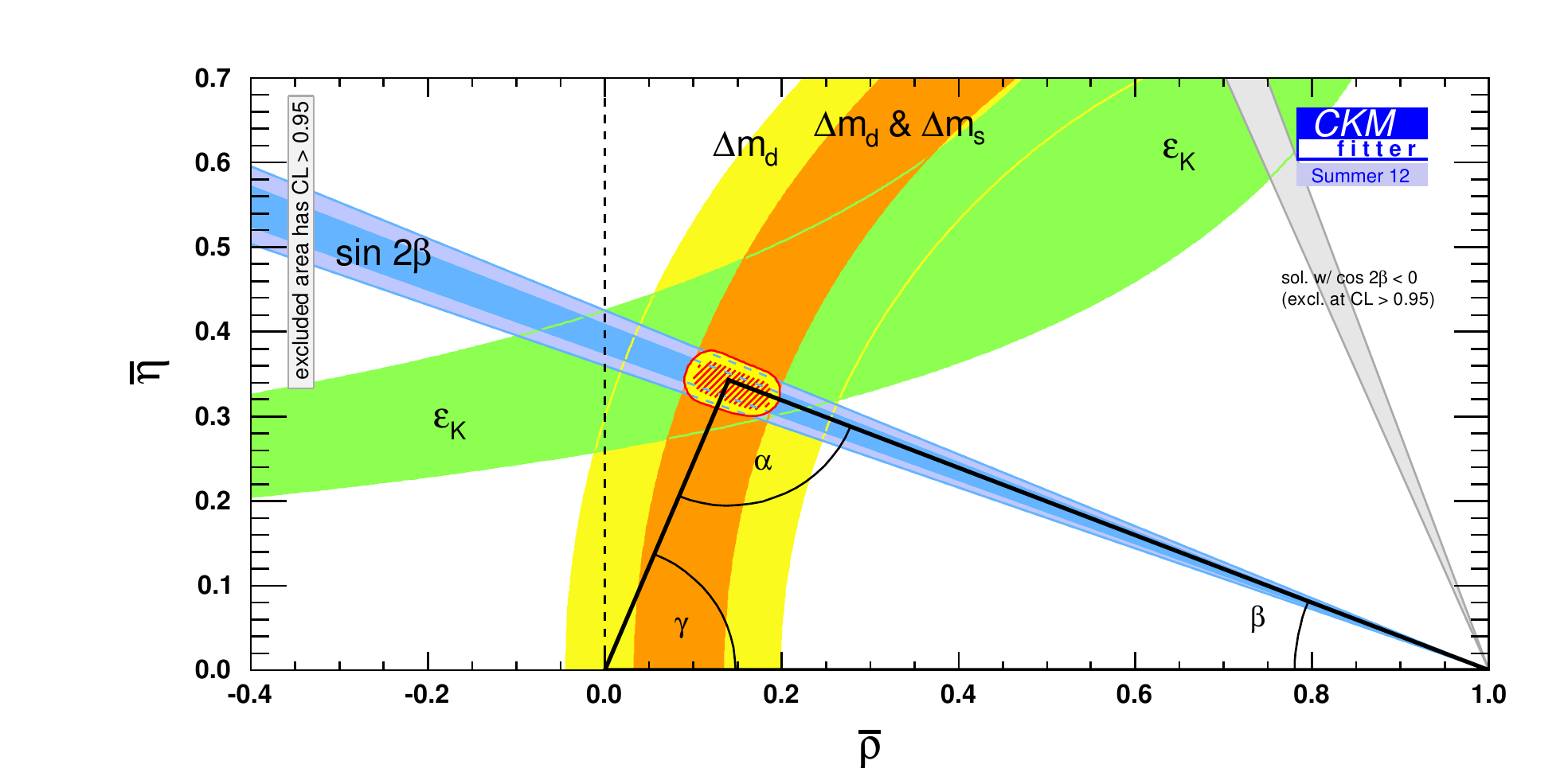}
\caption{\label{FigureCKMloop} Constraints from  ``Tree'' (left)  and   ``Loop'' (right)  quantities only~\cite{CKMfitter}.}
\end{figure*}

\begin{figure*}[htb]
\begin{center}
\includegraphics[width=.25\textwidth]{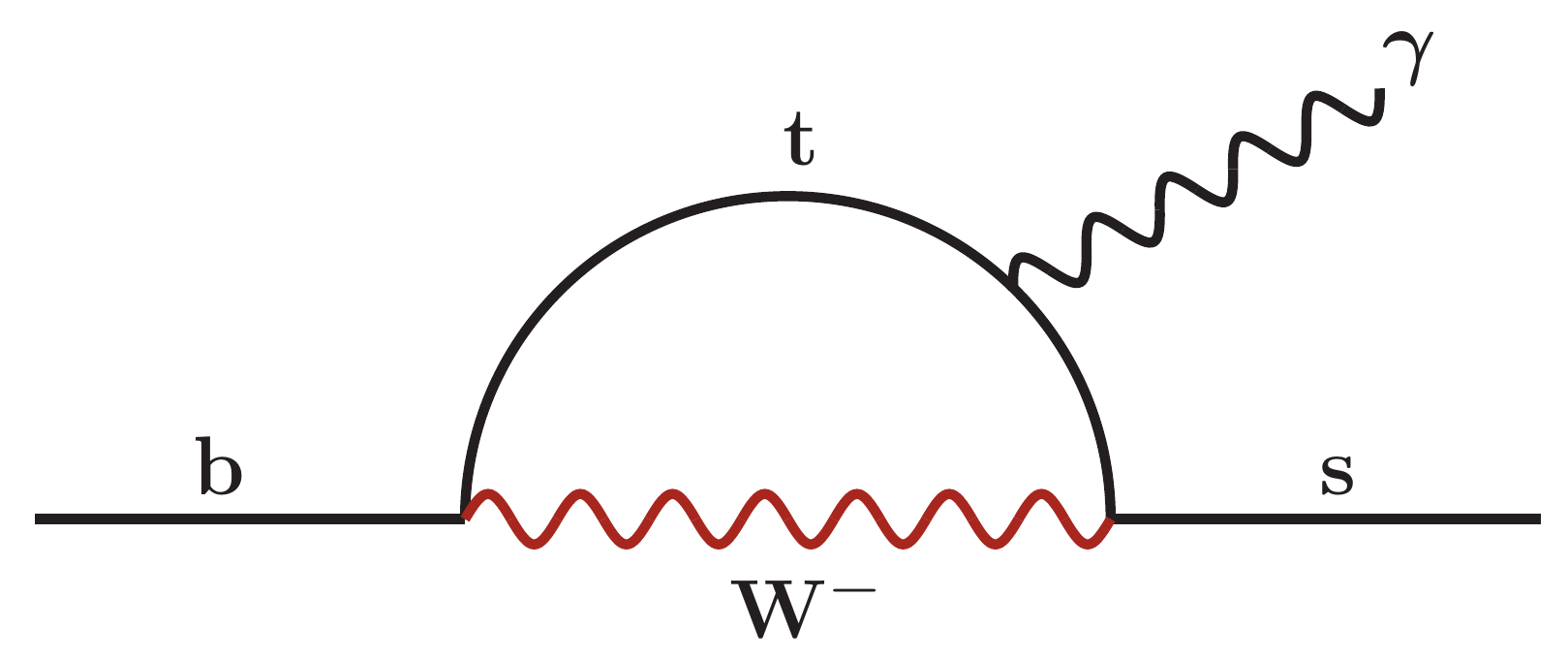}\quad\includegraphics[width=.25\textwidth]{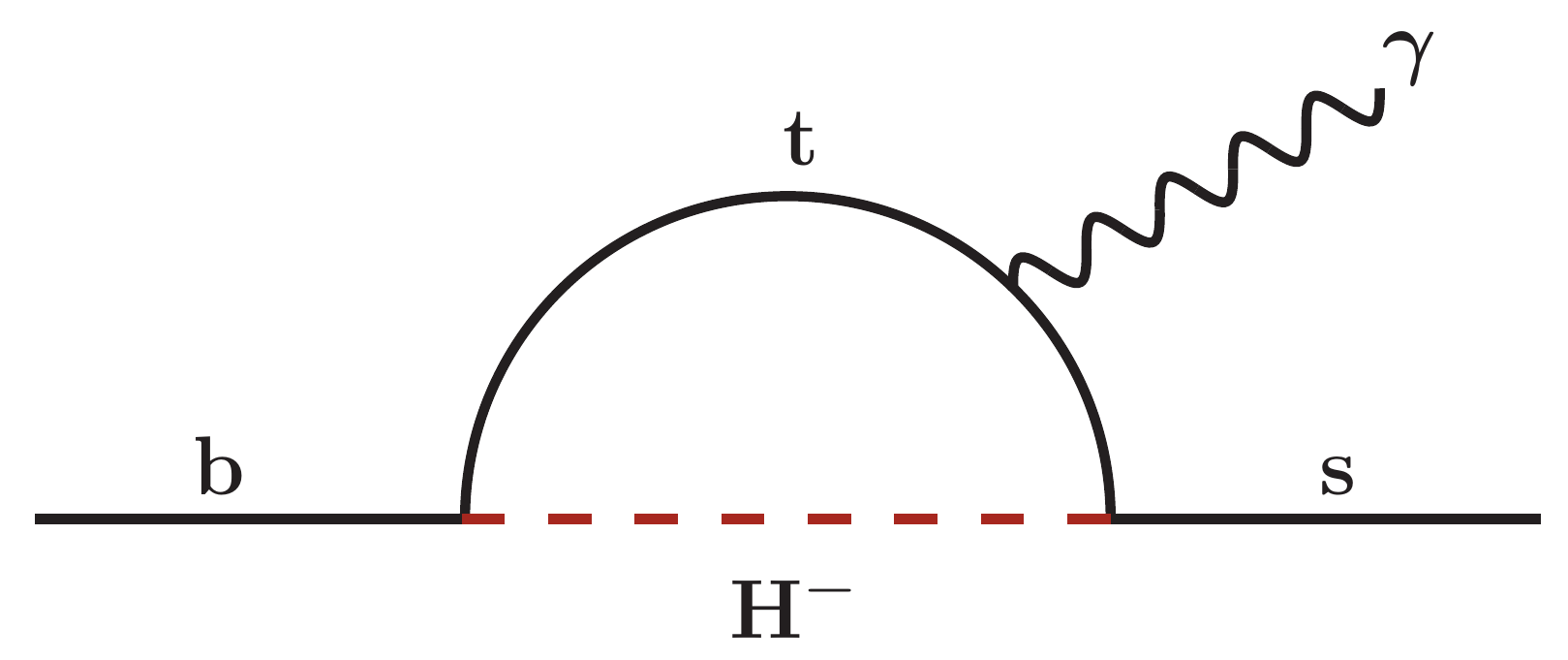}\quad\includegraphics[width=.25\textwidth]{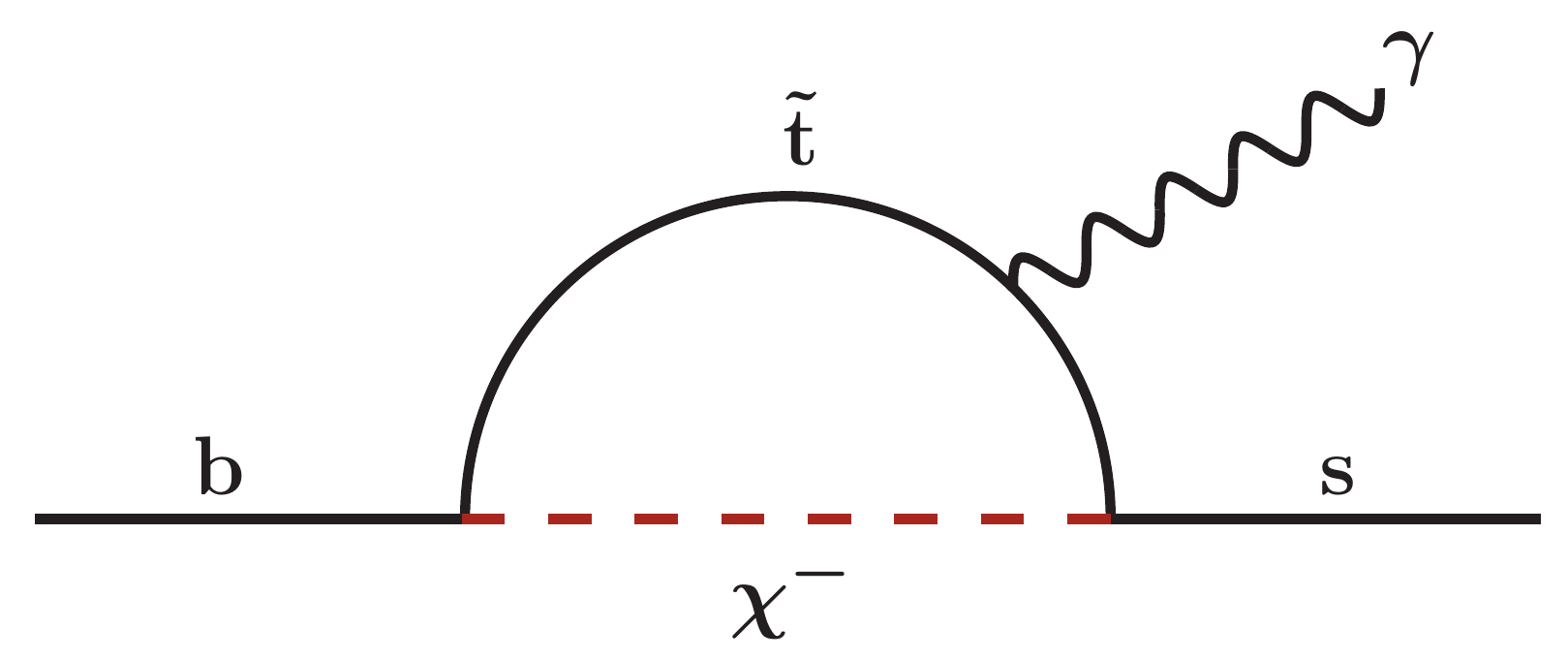}
\end{center}
\caption{Loop-induced $\bar B \to X_s \gamma$ decay via the SM particles, $W^-$ boson and top quark $t$ (left), via new particle, namely charged Higgs $H^-$ and top quark $t$ (middle), or via new supersymmetric particles, chargino $\tilde{\chi^-}$ and stop $\tilde{t}$ (right).}
\label{fig:feynmanbsgamma}
\end{figure*}

It can be illustrated by the over-constrained triangles in
 the complex plane which reflect the unitarity of the 
CKM matrix, see Figure~\ref{FigureCKM}. Some historical CKM fits in Figure~\ref{FigureCKMhistoric}\nocite{Ali:1994es,Harrison:1998yr}
illustrate the great success of the  $B$ factories. 
A closer look on the constraints is even more impressing: the constraints induced by CP conserving and by CP violating observables are fully consistent with 
each other (see Figure~\ref{FigureCKMCP2}).
Moreover, the tree-level observables which are in general assumed not being affected by new physics effects provide constraints which are fully consistent with 
the ones obtained from loop-induced observables (see Figure~\ref{FigureCKMloop}).
Especially this feature has been somehow unexpected because in principle
(loop-induced) flavour changing neutral current (FCNC) processes  like  $\bar B \to X_s  \gamma$
offer high sensitivity to new physics (NP)  due to the simple fact that additional contributions to the
decay rate, in which SM particles are replaced by new particles such as
the supersymmetric charginos or gluinos, are not suppressed by the
loop factor $\alpha/4\pi$ relative to the SM contribution, see Figure~\ref{fig:feynmanbsgamma}. 

It is worth mentioning that there is much more flavour data not shown in the unitarity fits which confirms the SM predictions 
of flavour mixing like rare decays.  This success of the CKM theory was honoured by the  Nobel Prize
in physics in 2008.

The absence of any unambiguous sign for NP in the flavour data but {\it also}  in the high-$p_T$ data of the ATLAS and CMS
experiments~(\citeauthor{ATLAS,CMS}) guides our attention to the well-known flavour problem of NP: 
{in the model-independent approach using 
the effective electroweak Hamiltonian, the contribution to one six-dimensional specific operator ${\cal O}_i$ can be parametrised via  
$( C^i_{\rm SM}\, / \,  (M_W)^2  +  C^i_{\rm NP}\, / \, (\Lambda_{\rm NP})^2 ) \times {\cal O}_i$
where the first term represents the SM contribution at the electroweak scale $M_W$ and the second one the NP 
contribution with an unknown coupling 
$C^i_{\rm NP}$ and an unknown  NP scale $\Lambda_{\rm NP}$.}  
The non-existence of large NP effects in  FCNC  observables in general
asks for an explanation why FCNC are suppressed. This famous flavour problem of NP can be solved in two ways: 
either the mass scale of the new degrees of freedom $\Lambda_{\rm NP}$  is very high or the 
new flavour-violating couplings $C^i_{\rm NP}$  are small for (symmetry?) reasons that remain to be found. 
For example,  assuming  {\it generic\/}  new flavour-violating couplings of $O(1)$,  
the present data on  $K$-$\bar K$ mixing implies  a very high NP scale of order $10^3$--$10^4$ TeV 
depending on whether the new  contributions enter at loop- or at tree-level.   
 In contrast, theoretical considerations on scale hierarchies in the Higgs sector, which is responsible for the mass \mbox{generation} 
of the fundamental particles in the SM, call for NP at
order $1$ TeV.    But 
any NP below the  $1$ TeV scale must have a non-generic flavour structure.

These considerations also imply that FCNC
decays provide information about the SM and its extensions via virtual
effects to scales presently not accessible by the direct search for new particles (for reviews see Refs.~\cite{Hurth:2010tk,Hurth:2003vb}). Thus,
the information offered by the FCNC is complementary to the one  provided by the  high-$p_T$  experiments ATLAS and CMS~\cite{Mahmoudi:2012uk, Hurth:2011zy}.
It is also obvious, that the indirect information on NP by FCNC (even if SM-like) will be most valuable when the general nature of NP will be identified 
in the direct search, especially when the mass scale of NP will be fixed.

Indeed, in the SM  the Glashow-Iliopoulos-Maiani (GIM)
mechanism, small CKM elements and often helicity,  all suppress FCNC
processes. These suppression factors stem from the particle content of
the SM  and the unexplained smallness of most Yukawa
couplings and are absent in generic extensions of the SM.  Hence FCNCs are an
excellent testing ground to probe new physics up to scales of 100 TeV,
depending on the model. Moreover, CP violation in flavour-changing transitions of
the SM  is governed by a single parameter, the phase of the
CKM matrix, so that the SM  is highly predictive about CP
physics.  Certain CP asymmetries are practically free of 
hadronic uncertainties, which permits the extraction of fundamental CP
phases from experiments with high accuracy. Thus,  CP physics is a powerful
tool to probe extensions of the SM, which generically involve 
many new CP phases.

As a consequence, the present data of the $B$ \mbox{physics} experiments  
already imply significant restrictions for the parameter 
space of new physics models  -- as we will explicitly show below --
and lead  to important 
clues for the direct search for new particles and for  model building 
beyond the SM. 

Thus,  the CKM mechanism is the dominating effect for CP
violation and flavour mixing in the quark sector; however, there is still room for sizeable new effects and new flavour structures because 
the flavour sector has only been tested  at the $10\%$ level especially in the $b-s$ sector. 
Moreover, the Standard Model does not describe the flavour phenomena in 
the lepton sector due to the existence of neutrino masses, a property not described by the SM. \\
Furthermore, while the gauge principle governs the gauge sector of the SM
there is no guiding principle in the flavour sector:
the CKM mechanism (three Yukawa SM couplings) provides a phenomenological 
description of quark flavour processes,  but leaves the significant hierarchy 
of the quark masses and the mixing parameters -- observed in experiment -- unexplained. 
This problem is often referred to as the flavour problem of the SM.

{There are many solutions to this problem proposed in the literature, for example the Froggatt-Nielsen mechanism~\cite{Froggatt:1978nt}
and  the Nelson-Strassler mechanism~\cite{Nelson:2000sn};
the popular Randall-Sundrum model is another approach  to this  SM flavour problem, where the hierarchy  of the flavour parameters 
can be explained by the special geometrical settings of the model. In addition the so-called gauge-hierarchy problem
in the Higgs sector finds a natural  explanation in this model~\cite{Randall:1999ee,Grossman:1999ra,Gherghetta:2000qt}.}

The SM flavour problem  is also reflected in the fact that 
many open fundamental questions of particle physics are related 
to flavour:
\begin{itemize}
\item How many families of fundamental fermions are there?
\item How are neutrino and quark masses and mixing angles generated?
\item Do there exist new sources of flavour and CP violation?
\item Is there CP violation in the QCD gauge sector?
\item Are there relations between the flavour structure in the lepton and quark sectors?   
\end{itemize}

There is  already experimental evidence beyond the SM which
is partially  connected to flavour physics:  the existence of dark
matter, the non-zero neutrino masses, and the baryon asymmetry of the 
universe; the latter  implies the need for new sources of CP violation beyond
 the one offered  by the SM. This provides an important link between particle 
 physics  and cosmology.

In the following sections, we discuss the latest key measurements  by LHCb, the $B$ factories, and the Tevatron experiments.

\section{Latest measurements at hadron colliders}
\subsection{New physics in $\boldsymbol{B_q -\bar B_q}$ mixing ($\boldsymbol{q=d,s}$)?}
\begin{figure}[!t]
\begin{center}
\begin{minipage}{12pc}
\includegraphics[width=10pc]{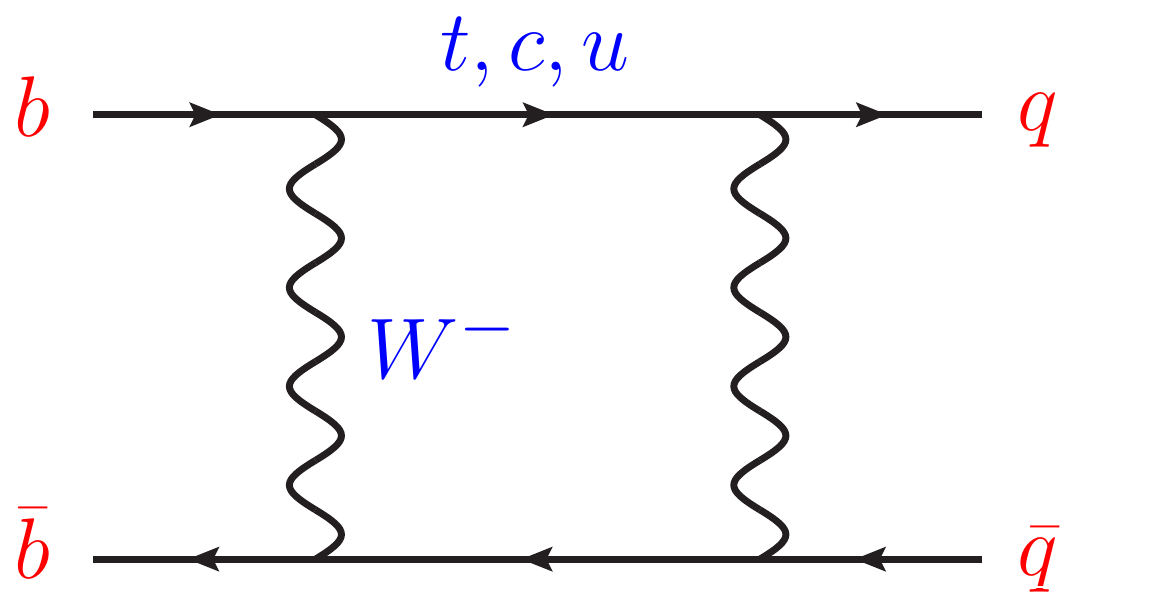}
\end{minipage}
\caption{$B_q -\bar{B_q}$ mixing governed by the box diagram.} \label{boxdiagram} 
\end{center}
\end{figure}
The meson-antimeson oscillation is governed by two parameters, the mass difference ($\Delta M$) of the two physical eigen states $B_H$ and $B_L$ and the  decay rate difference ($\Delta \Gamma$):
\begin{eqnarray}
\Delta M &:=& M_H - M_L = 2 |M_{12}|\,,\\
\Delta \Gamma &:=& \Gamma_L - \Gamma_H = 2 |\Gamma_{12}| \cos \Phi\, . 
\end{eqnarray}
$|M_{12}|$  corresponds to the dispersive part of the box diagram in Figure~\ref{boxdiagram}  which is sensitive to new heavy particles while $\Gamma_{12}$  corresponds to its  absorptive part which is sensitive to the light internal particles, and, thus, often assumed to be insensitive to NP. 
Possible NP effects can be parametrised by the  complex parameter $\Delta_q \, (q = d, s)$:  $M_{12,q} = M_{12,q}^{SM} \times \Delta_q$ in a model-independent way. There are several
observables which are sensitive to the NP phase $arg(\Delta_q) =  \Phi_q^\Delta$, for example  $\Delta  M_q$ and $\Delta |\Gamma_q| =  2 |\Gamma_{12,q}| \times \cos(\Phi_q^{SM} + \Phi_q^\Delta)$. But also the golden modes   $B_d \to J/\psi K_s^0$ and  $B_s \to J/\psi \Phi$ are sensitive to the NP phases.  The corresponding
CP violating phases in the SM $\beta_q^{SM}$ are modified via  $2 \beta_d^{SM} + \Phi_d^\Delta$ and $2 \beta_s^{SM}  - \Phi_s^\Delta$. 
\begin{figure}[!t]
\begin{center}
\begin{minipage}{20pc}
\includegraphics[width=20pc]{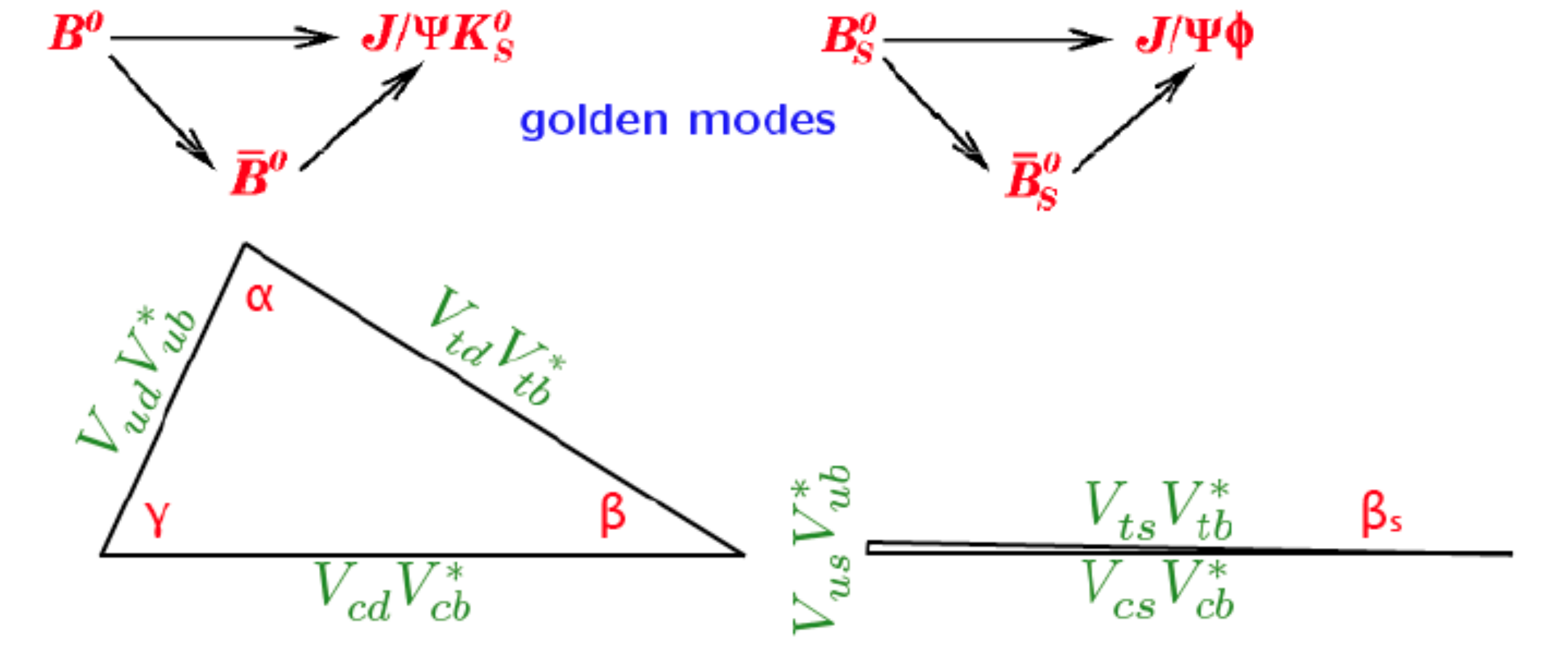}
\end{minipage}
\caption{CP violating through interference  of decay with and without mixing in the two golden modes of the $B_d$ and $B_s$ system.} \label{betas}
\end{center}
\end{figure}

\begin{figure}[!t]
\begin{center}
\includegraphics[width=7cm]{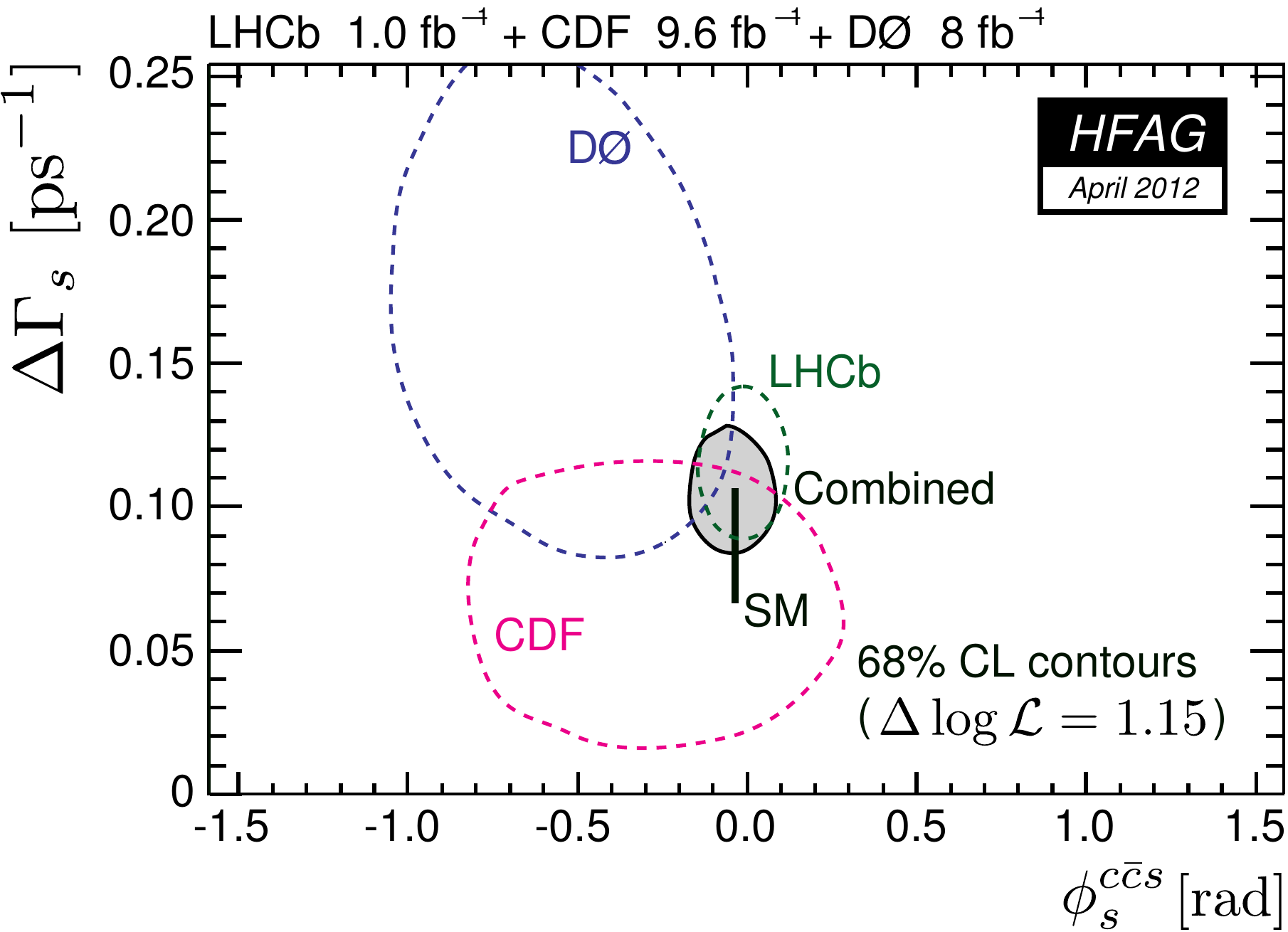}
\caption{HFAG 2012 combination of $\phi_s$ and $\Delta \Gamma_s$ results \cite{Amhis:2012bh}.}
  \label{Figure:mixing}
\end{center}
\end{figure}

As illustrated in Figure~\ref{betas}, the CP violating phase in $B_s \to J/\psi \Phi$ is very small in the SM{~\cite{Lenz:2010gu}: 
\begin{equation}
2 \beta_s^{SM}  = - {\rm arg}\left((V_{ts} V_{tb}^*)^2/(V_{cs} V^*_{cb})^2\right)  = (2.1 \pm 0.1)^0. 
\end{equation}
}%
LHCb has reported a measurement of this small angle~\cite{LHCb:2011aa}  which is fully consistent  with  the SM prediction and also consistent with the previous measurements of CDF and D0. In addition, LHCb  has resolved the two-fold ambiguity~\cite{Aaij:2012eq} and reported their first measurement of $\Delta \Gamma_s$ which confirms the Heavy Quark Expansion prediction:
\begin{eqnarray}
\Delta \Gamma_s (\mbox{LHCb}) =& ( 0.116 \pm 0.019) \mbox{ ps}^{-1},&~\mbox{\cite{LHCb:2011aa}}\\
\Delta \Gamma_s (\mbox{HFAG}) =& ( 0.105 \pm 0.015) \mbox{ ps}^{-1},&~\mbox{\cite{Amhis:2012bh}}\\
\Delta \Gamma_s (\mbox{SM}) =& ( 0.087 \pm 0.021) \mbox{ ps}^{-1},&~\mbox{\cite{Lenz:2011ti}}
\end{eqnarray}
and
\begin{eqnarray}
\phi_s (\mbox{LHCb}) =& (-0.001 \pm 0.104) \mbox{ rad},&~\mbox{\cite{LHCb:2011aa}}\\
\phi_s (\mbox{HFAG}) =& (-0.044^{+0.090}_{-0.085}) \mbox{ rad},&~\mbox{\cite{Amhis:2012bh}}\\
\phi_s (\mbox{SM}) =& (-0.036 \pm 0.002) \mbox{ rad}.&~\mbox{\cite{Lenz:2011ti}}
\end{eqnarray}
Thus, NP contributions in the mixing of the $B_s$ system are disfavoured by the present data (see Figure~\ref{Figure:mixing}).   

Furthermore, the semi-leptonic asymmetries offer an independent test of NP physics in $B_q - \bar B_q$ mixing. In the presence of NP 
they get modified via 
\begin{equation}
a_{sl}^q = {\rm Im}\left(\frac{\Gamma_{12,q}}{M_{12,q}}\right) = \left(\frac{|\Gamma_{12,q}|}{  |M_{12,q}|}\right) \, \frac{\sin(\Phi_q^{SM}  + \Phi_q^{\Delta})}{|\Delta_q|}.
\end{equation}
D0 had measured the dimuon charge  asymmetry to disagree with the SM prediction by 3.9$\sigma$ \cite{Abazov:2011yk,Lenz:2011ti}:
\begin{eqnarray}
A_{sl}^b (\mbox{D0})& = & -(7.87 \pm 1.72 \pm 0.93) \times 10^{-3},\\
A_{sl}^b (\mbox{SM})& = & -(0.28^{+0.05}_{-0.06}) \times 10^{-3},
\end{eqnarray}
where $A_{sl}^b$ is a linear combination of the semi-leptonic asymmetries $a_{sl}^d$ and $a_{sl}^s$. 
As was argued in Ref.~\cite{Lenz:2011ww} , the central value of the D0 measurement is larger than theoretically possible.   
\begin{figure*}[!t]
\begin{center}
\includegraphics[width=7.5cm]{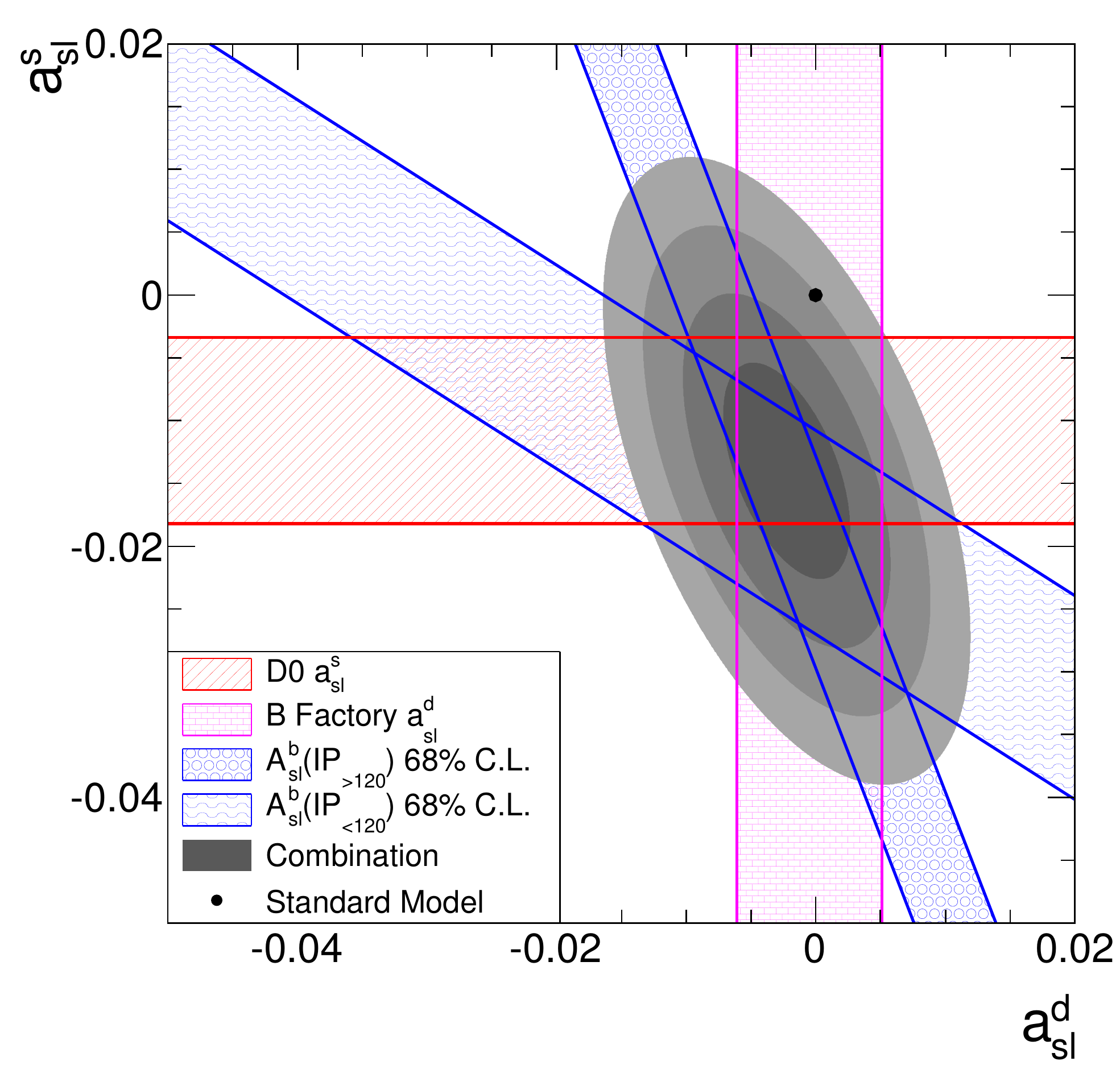}\quad\quad\raisebox{0.3cm}{\includegraphics[width=7.5cm,height=6.8cm]{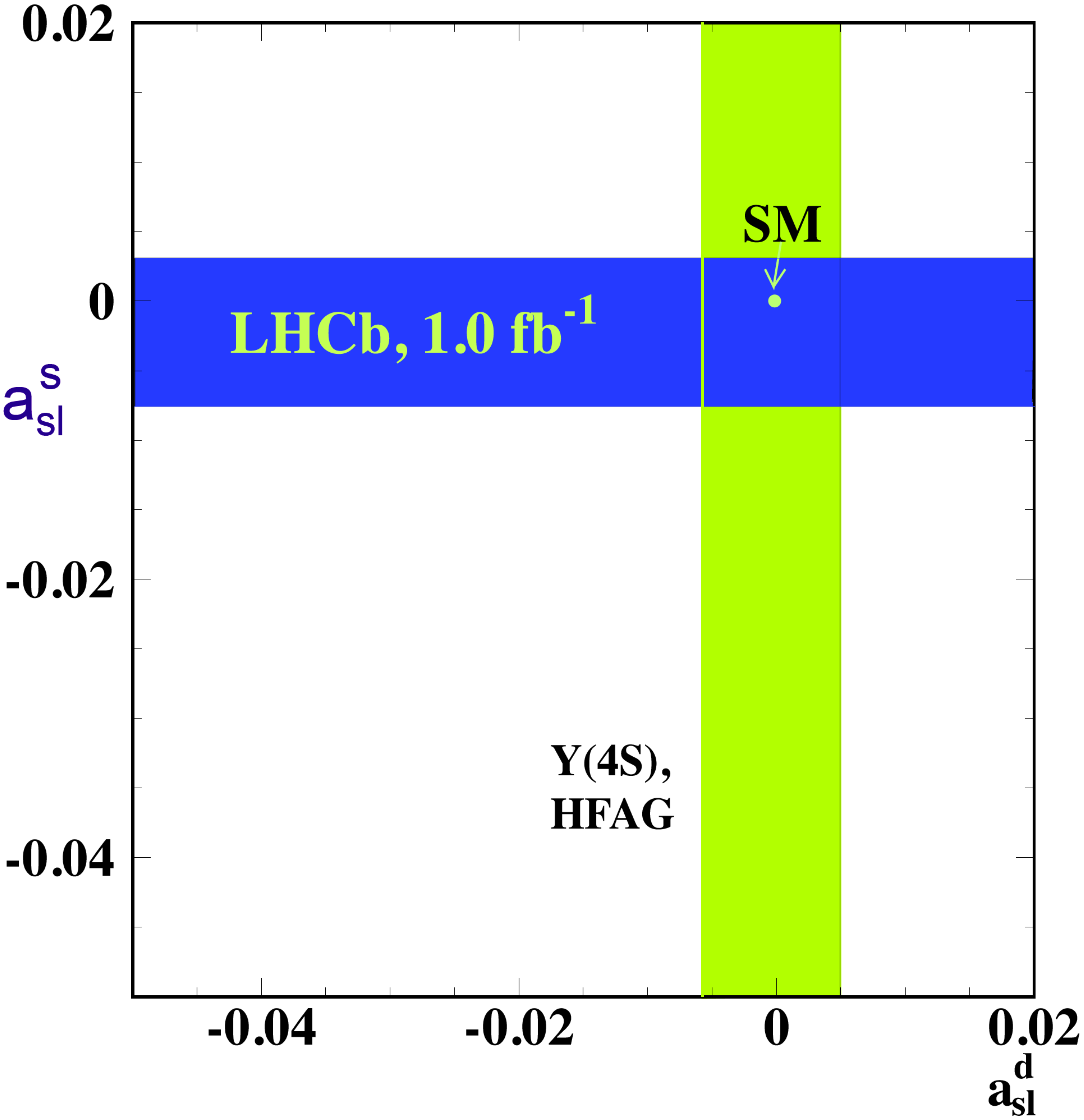}}\vspace*{-0.2cm}
\caption{Measurements of the semileptonic CP asymmetries $a_{sl}^{d}$ and $a_{sl}^s$ by D0 (left,~\cite{Abazov:2012zz})  and by LHCb and $B$ 
factories (right,~\cite{LHCb-CONF-2012-022}).}
  \label{semileptonic}
\end{center}
\end{figure*}
\begin{figure*}[!t]
\begin{center}
\includegraphics[width=7.8cm]{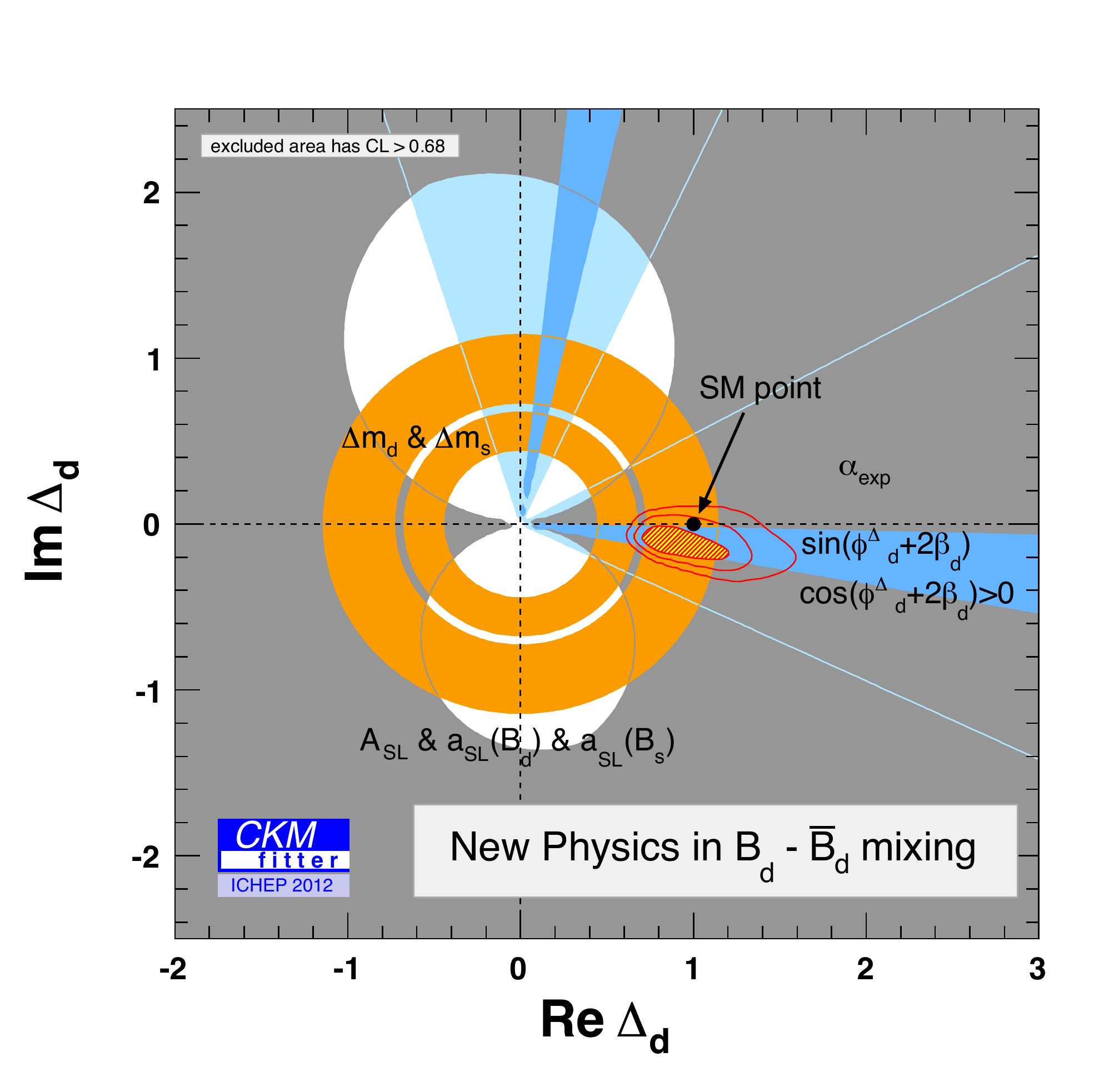}\quad\includegraphics[width=7.8cm]{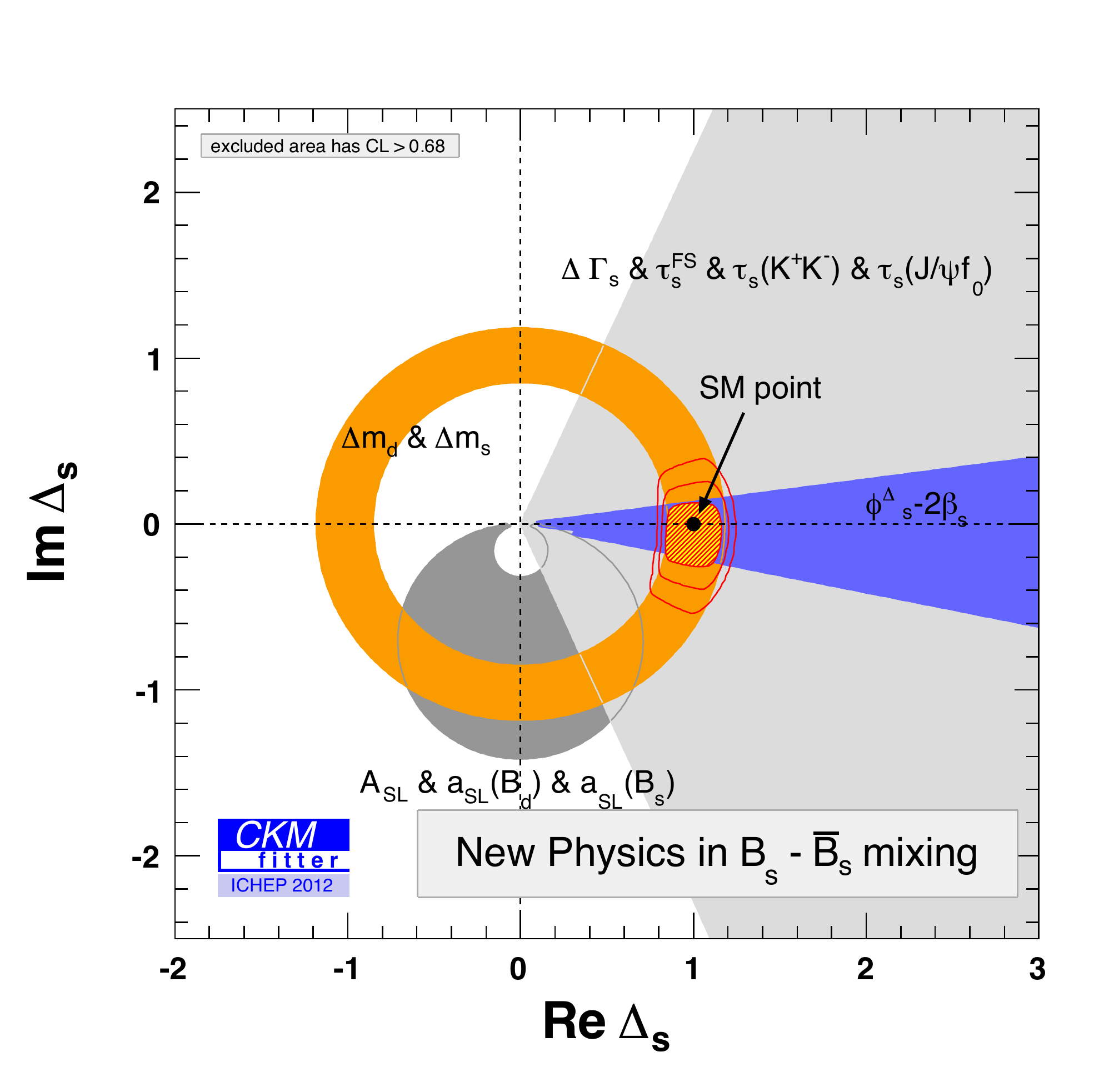}\vspace*{-0.2cm}
\caption{New physics in $B_d - \bar B_d$ (left) and in $B_s - \bar B_s$ (right) mixing: fit result for the complex parameters $\Delta_d$ and $\Delta_s$, respectively. 
The (red) hashed  area shows the region with C.L. $< 68.3\%$ while the two additional contour lines inscribe the regions with C.L. $< 95.45\%$ and C.L. $< 99.73\%$, respectively~\cite{CKMfitter}.}
  \label{Bsdmixing}
\end{center}
\end{figure*}
More recently, there are also  direct measurements  of $a_{sl}^s$ and $a_{sl}^d$ by D0~\cite{Abazov:2012zz} which in combination with the dimuon charge  asymmetry still lead to a $3 \sigma$ deviation from the SM prediction, see left plot of Figure~\ref{semileptonic}.
In contrast, the first LHCb measurement of $a_{sl}^s$~\cite{LHCb-CONF-2012-022} and the measurement of $a_{sl}^d$ by the $B$ factories~\cite{Amhis:2012bh} 
are nicely compatible  with the SM predictions, see right plot of Figure~\ref{semileptonic}.   
Obviously, there is a slight  tension between the two data sets which calls for improved measurements. 

Finally, we mention that within the model-independent analysis of NP in $B_d- \bar B_d$ mixing, a $1.6 \sigma$  deviation is obtained for the 2-dimensional SM hypothesis 
$\Delta_d =1$.  Figure~\ref{Bsdmixing} shows the fit result for the complex parameter $\Delta_d$.   It is worth mentioning that a NP phase $\Phi_d^\Delta < 0$ would resolve the slight tension between ${\mathrm{BR}}(B\to \tau\nu)$ and $\sin\beta$ in the global
CKM fit (see subsection~\ref{Btaunu}).  We also state that in the $B_s$ system the CKMfitter group finds a $0.2 \sigma$ deviation for the corresponding SM hypothesis 
$\Delta_s=1$, see Figure~\ref{Bsdmixing}.   A  detailed discussion can be found in \cite{Lenz:2012az}.

\subsection{Angular observables in $\boldsymbol{B \to K^* \ell^+\ell^-}$}
The semi-leptonic decay $B \to K^* \ell^+\ell^-$ is mediated by electroweak loop diagrams in the SM and can receive large enhancements from NP. It gives access to a variety of angular observables and hence offers a rich phenomenology. From the theoretical point of view, exclusive modes suffer  from large hadronic uncertainties due to the form factors. One has to find strategies to reduce this form factor dependence by considering appropriate ratios. 
On the contrary, the experimental measurements are easier here as compared to the case of  inclusive modes.

Two kinematic regimes are considered in order to avoid the narrow $c \bar c$ resonances. In the region where the dimuon invariant mass squared, $q^2$, is small ($1 < q^2 < 6$ GeV$^2$) the decay is described by the QCD-improved Factorisation (QCDF) and the Soft-Collinear Effective Theory (SCET). In the high $q^2$ region ($q^2 \gtrsim 14$ GeV$^2$) on the other hand the Operator Product Expansion (OPE) is used. As the theoretical treatments in the low- and high-$q^2$ regions are based on different concepts, the consistency of the consequences from the two regimes allows for important cross checks.   
\begin{figure}[!t]
\vspace*{0.1cm}\includegraphics[width=8.7cm]{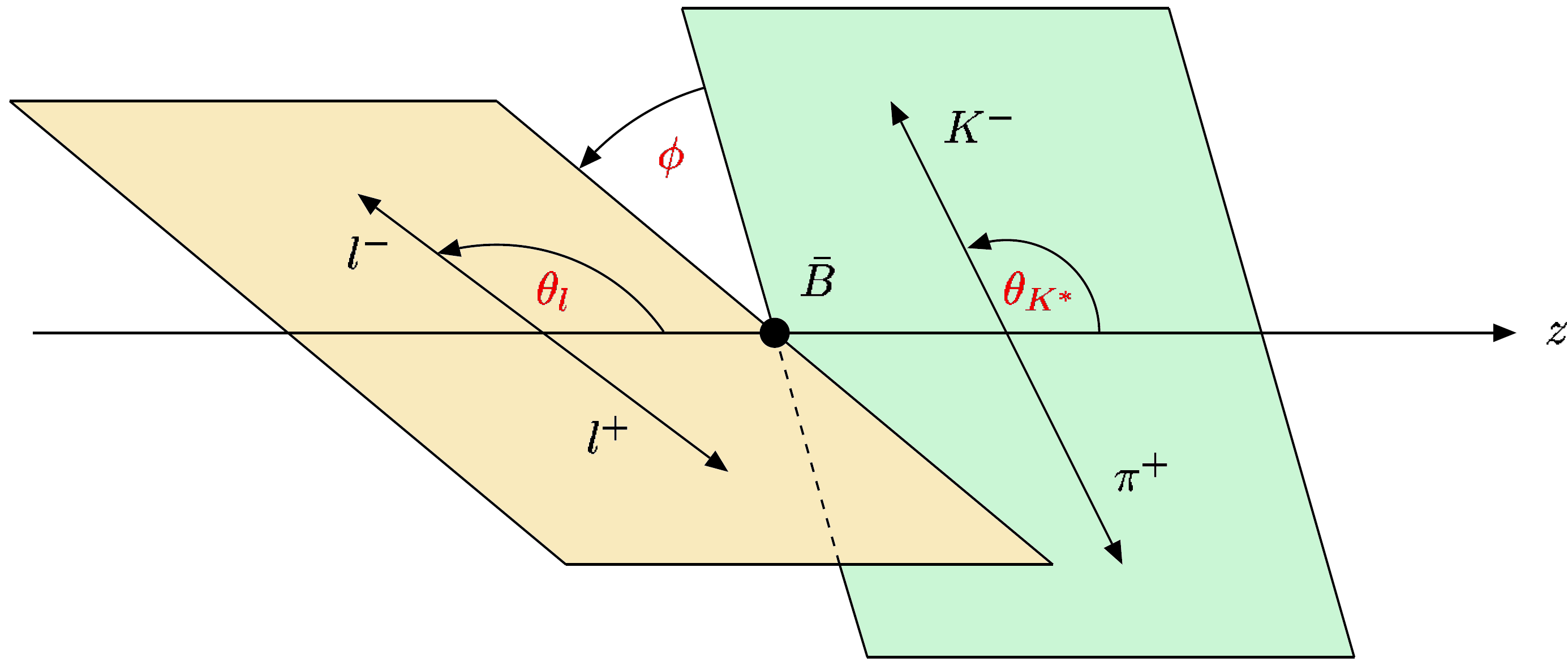}
\caption{Kinematic variables in $B \to K^* \ell^+\ell^-$.}
  \label{Figure:bsll_ang}
\end{figure}
\begin{figure*}[!t]
\begin{center}
\includegraphics[width=8.3cm,height=5.cm]{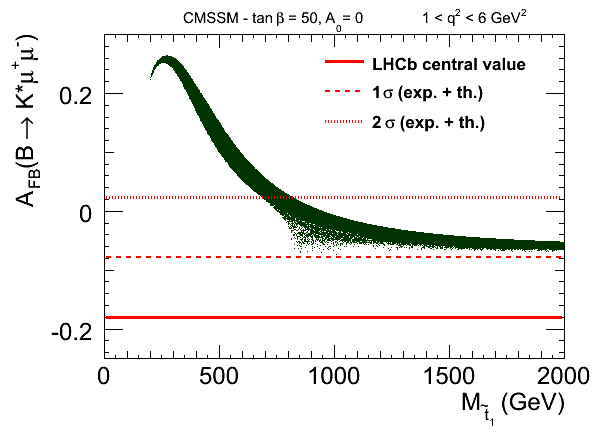}\includegraphics[width=8.3cm,height=5.cm]{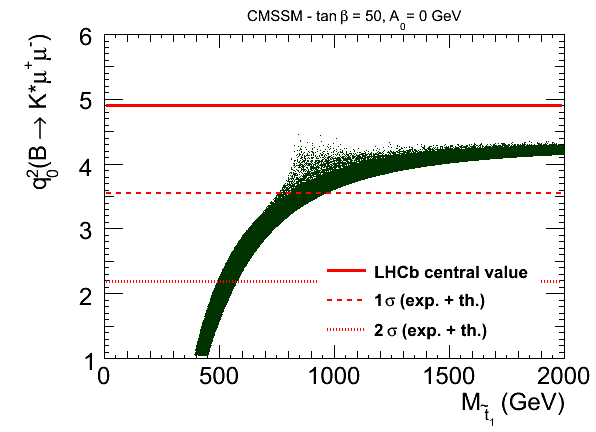}
\caption{SUSY spread of $A_{FB}$ (left) and the $A_{FB}$ zero-crossing, $q_0^2$ (right) as a function of the lightest stop mass in the CMSSM for $\tan\beta= 50$ and $A_0=0$.}
  \label{Figure:AFB}
\end{center}
\end{figure*}
\begin{figure*}[!t]
\begin{center}
\includegraphics[width=7.5cm]{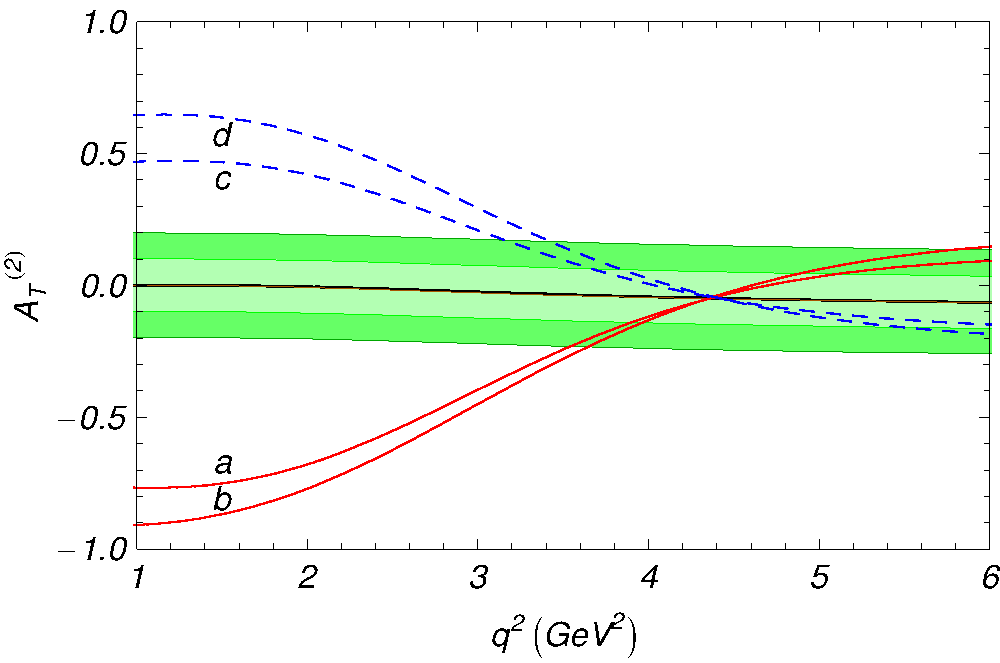}\quad\quad\includegraphics[width=7.5cm]{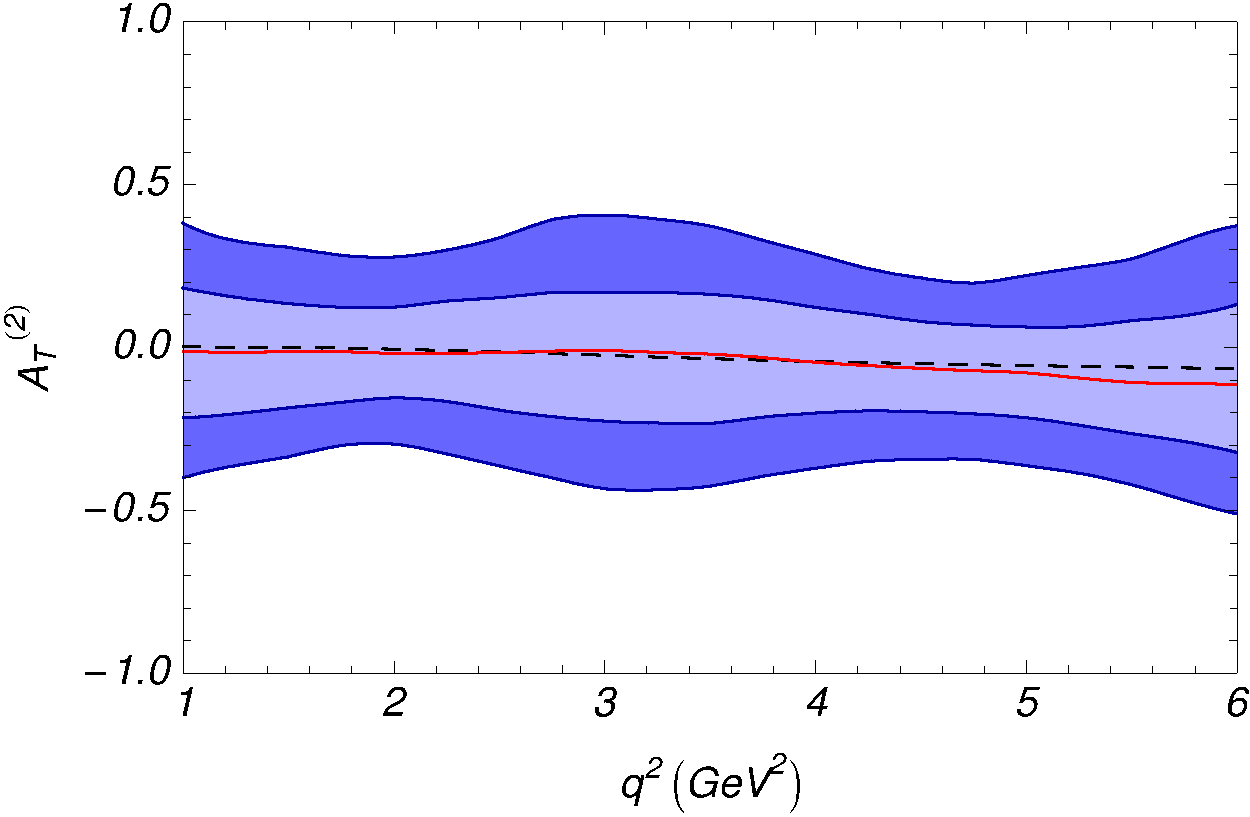}
\caption{The theoretical errors (left) for $A_T^{(2)}$ are compared to the experimental errors (right) as a function of $q^2$. Light (green) bands include an estimated $\Lambda/m_b$ uncertainty at a $\pm 5\%$ level and the dark (green) bands correspond to a $\pm 10\%$ correction. The curves labelled (a)--(d) correspond to different benchmark SUSY scenarios \cite{Egede:2008uy}. In the right plot, the light and dark (blue) bands correspond to 1$\sigma$ and 2$\sigma$ statistical errors with a yield corresponding to 10 fb$^{-1}$ data from LHCb, respectively.}
\label{Figure:AT2}
\end{center}
\end{figure*}

The angular distribution of $B \to K^* \ell^+\ell^-$ with $K^* \to K^+ \pi^-$ can be fully described in terms of four kinematic variables, the angles $\theta_\ell, \theta_K, \phi$ and $q^2$ as shown in Figure~\ref{Figure:bsll_ang}. There are twelve angular terms appearing in the differential decay rate that can be exploited experimentally. The full expressions for these functions can be found in \cite{Egede:2008uy,Egede:2010zc}.

Several angular observables, namely the differential branching ratio, forward-backward asymmetry ($A_{FB}$) and $K^*$ longitudinal fraction ($F_L$), have already been measured by the Belle and BaBar experiments, and also CDF and LHCb. In addition, LHCb has also measured $S_3$ which is related to the asymmetry between the $K^*$ parallel and perpendicular spin amplitudes, and the value of $q_0^2$ for which the differential forward-backward asymmetry vanishes. The experimental results as well as the SM predictions for these observables are summarised in Table~\ref{Inputobservables}. They agree within the current errors. 

In the Constrained MSSM (CMSSM), $A_{FB}$ and $q_0^2$ are particularly constraining.
The CMSSM is governed by only five additional universal parameters {defined at the $M_{\rm GUT}$ scale}: the mass of the scalar particles, $m_0$, the mass of the gauginos, $m_{1/2}$, the trilinear coupling, $A_0$, the ratio of the vacuum expectation values of the Higgs doublet, $\tan\beta$, and finally the sign of the higgsino mass term, $\mu$.  
In Figure~\ref{Figure:AFB}   the SUSY spread is compared to the LHCb 1 and 2$\sigma$ bounds in the CMSSM parameter space with $\tan\beta= 50$ and $A_0=0$ \cite{Mahmoudi:2012un}.

With 2--3 fb$^{-1}$ of integrated luminosity, LHCb will have the opportunity of performing a full angular analysis. This calls in turn for optimised set of observables with reduced theoretical uncertainty. In particular, as the amplitudes depend linearly on the soft form factors at leading order in the low-$q^2$ region, a complete cancellation of the hadronic uncertainties could be possible in leading order, which consequently increases the sensitivity to new physics. In the high-$q^2$ {region, there} are improved Isgur-Wise relations between the form factors which allow
to construct optimal observables. 

Examples of such observables are the transversity amplitudes, $A_T^{(2,3,4,5)}$ \cite{Egede:2008uy,Egede:2010zc} (or similarly $P_{1...6}$~\cite{Matias:2012xw} and $H_T^{(1,2,3)}$ \cite{Bobeth:2010wg}). The sensitivity of $A_T^{(2)}$ to NP scenarios is illustrated in Figure~\ref{Figure:AT2}.  There exist a large number of analyses on the NP sensitivity showing the 
rich phenomenology of the angular observables~\cite{Egede:2008uy,Egede:2010zc,Matias:2012xw,Mahmoudi:2012un,Bobeth:2010wg,Altmannshofer:2008dz,Bobeth:2008ij,Bobeth:2011gi,Beaujean:2012uj}.

\begin{figure*}[!t]
\hspace{1cm}\includegraphics[height=5pc]{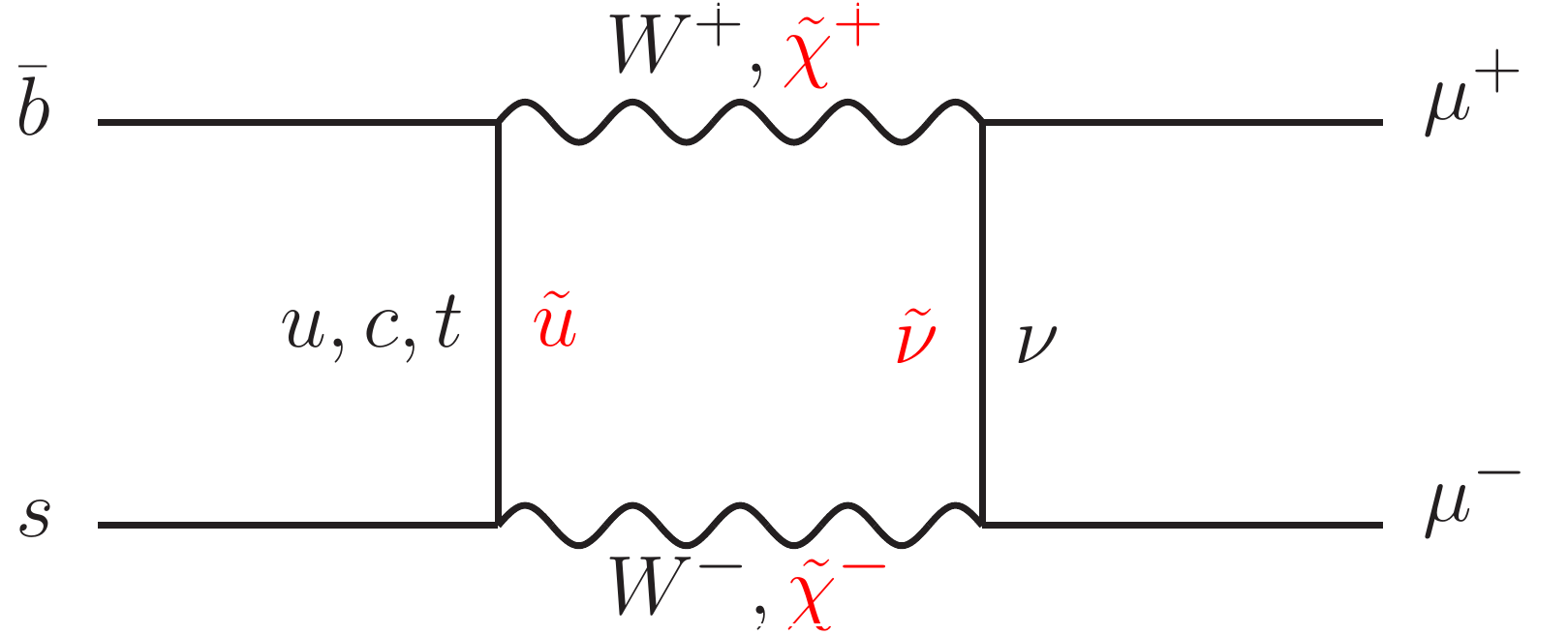}\quad\raisebox{3mm}{\includegraphics[height=4pc]{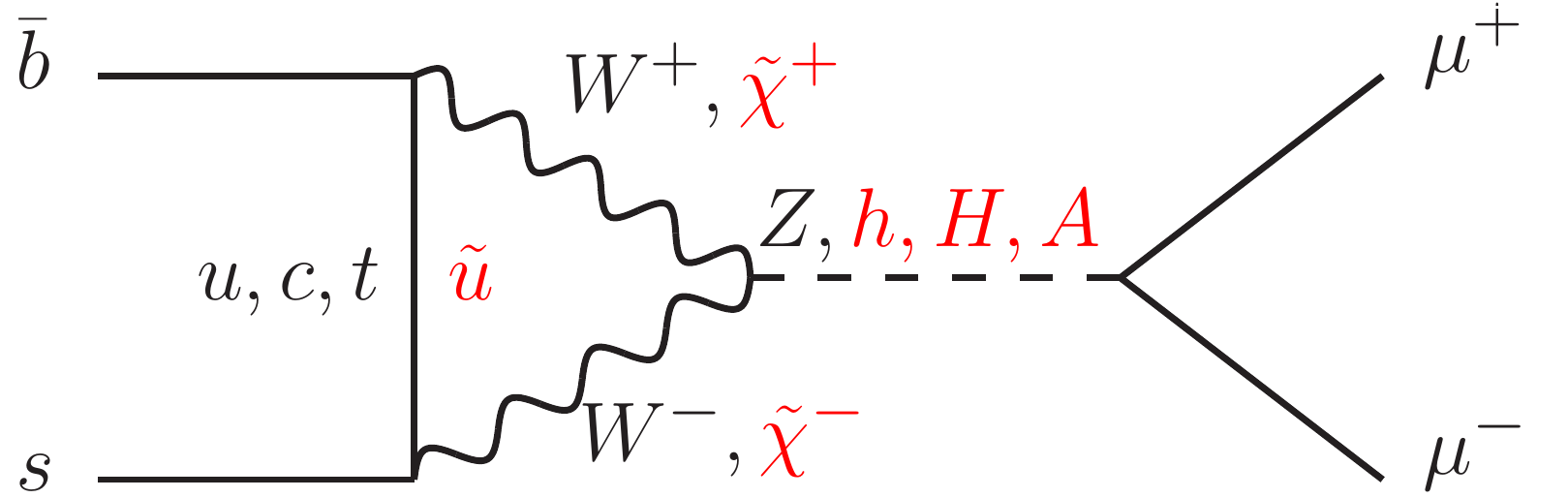}}
\caption{\label{Figure:Bsmumu} Contributions to the rare decay  $B_s \to \mu^+ \mu^-$  in the SM (black) and in the MSSM (light, red).}
\end{figure*}

\subsection{Implications of the latest measurements of $\boldsymbol{B_s \to \mu\mu}$}
The rare decay $B_s \to \mu^+ \mu^-$ proceeds via $Z^0$ penguin and box diagrams in the SM, see Figure~\ref{Figure:Bsmumu}. It is highly helicity-suppressed by 
a suppression factor $m_\mu/m_b$ on the amplitude level. As a consequence the SM prediction for the branching ratio of the decay $B_s \rightarrow \mu^+ \mu^-$
is of order $10^{-9}$.  However, the branching ratio  can be much
larger within specific extensions of the SM. 
For example, the helicity-suppression of the SM contribution
leads to an enhanced sensitivity to the Higgs-mediated 
scalar FCNCs within the 2HDM and, especially  within the MSSM, see Figure~\ref{Figure:Bsmumu}.
These non-standard contributions lead to a drastic 
enhancement in the large $\tan \beta$-limit~\cite{Huang:1998vb,Hamzaoui:1998nu,Babu:1999hn}. 
In the MSSM there is an enhancement factor of $(\tan\beta)^3$ on the amplitude level. 
The best upper limit for $\mathrm{BR}(B_s\to\mu^+\mu^-)$ measured in a single experiment comes from LHCb~\cite{Aaij:2012ac}:
\begin{equation}
\mathrm{BR}(B_s\to\mu^+\mu^-) < 4.5 \times 10^{-9}
\label{bsmumu}
\end{equation}
at 95\% C.L. This upper limit is followed by the result from CMS, $\mathrm{BR}(B_s\to\mu^+\mu^-) < 7.7 \times 10^{-9}$~\cite{Chatrchyan:2012rg}. 
The CDF collaboration obtains a 95\% C.L. upper limit, $\mathrm{BR}(B_s\to\mu^+\mu^-) < 3.4 \times 10^{-8}$~\cite{Aaltonen:2011fi}, together with a one sigma interval, $\mathrm{BR}(B_s\to\mu^+\mu^-)  = (1.3^{+0.9}_{-0.7})\times10^{-8}$, coming from an observed excess over the expected background. The ATLAS collaboration has announced the  upper limit,
$\mathrm{BR}(B_s\to\mu^+\mu^-) < 2.2  \times 10^{-8}$~\cite{Aad:2012pn}.
The combination of LHCb, ATLAS and CMS results leads to an upper bound of $4.2 \times 10^{-9}$ \cite{LHCb-CONF-2012-017}.

{Very recently, the LHCb collaboration announced the first evidence for the decay $\mathrm{BR}(B_s\to\mu^+\mu^-)$ with the branching ratio~\cite{Aaij:2012ct}:
\begin{equation}
\mathrm{BR}(B_s\to\mu^+\mu^-) = \left(3.2^{+1.4}_{-1.2}({\rm stat})^{+0.5}_{-0.3}({\rm syst})\right)\times 10^{-9}.
\label{bsmumu_evidence}
\end{equation}
This new measurement is a major step which will hopefully be followed by more precise results. The present accuracy however does not lead to improved constraints on supersymmetry as compared to the one from the previous upper limit. Nevertheless, as we will see later, the lower bound has consequences on the constraints on the Wilson coefficients in the MFV framework.
}

All these {results are} very close to the SM prediction, which is $\mathrm{BR}(B_s\to\mu^+\mu^-) =(3.53 \pm  0.38)  \times 10^{-9}$ \cite{Mahmoudi:2012un}.
The main theoretical uncertainty comes from the $B_s$ decay constant, which is now in the focus of the lattice gauge theory community, see Refs. \cite{Dimopoulos:2011gx,Bazavov:2011aa,Neil:2011ku,Na:2012kp,McNeile:2011ng,Davies:2012qf}.

The theoretical prediction does not directly correspond to the experimental branching ratio. There are two correction factors of $O(10\%)$: one includes the effect of the $\bar B_s - B_s$ oscillation~\cite{DeBruyn:2012wj,DeBruyn:2012wk}, the other takes into account effects of soft radiation~\cite{Buras:2012ru}. 

\begin{figure*}[!t]
\begin{center}
\includegraphics[width=7.5cm]{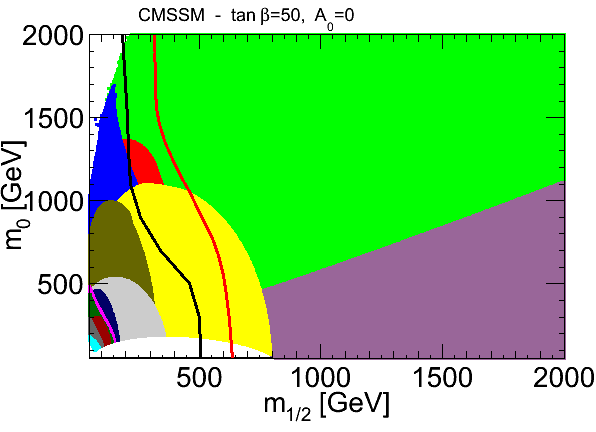}
\includegraphics[width=7.5cm]{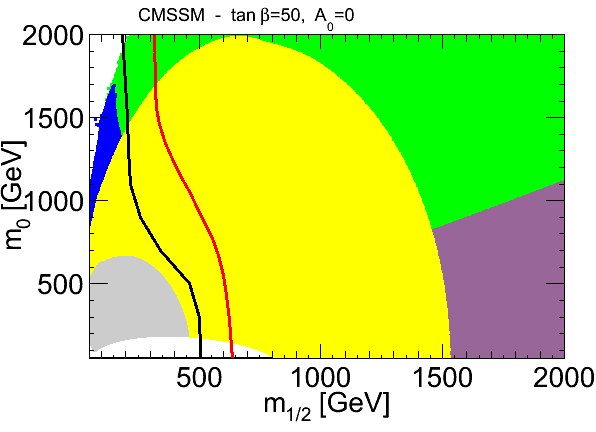}
\caption{Constraints from flavour observables on the  CMSSM in the plane ($m_{1/2}, m_0$) for $\tan\beta= 50$ with 2010 results on BR($B_s\to \mu^+\mu^-$) (left) and with the 2011 results (right).}
\label{Figure:Nazila1}
\end{center}
\end{figure*}
\begin{figure*}[!t]
\begin{center}
\includegraphics[width=7.5cm]{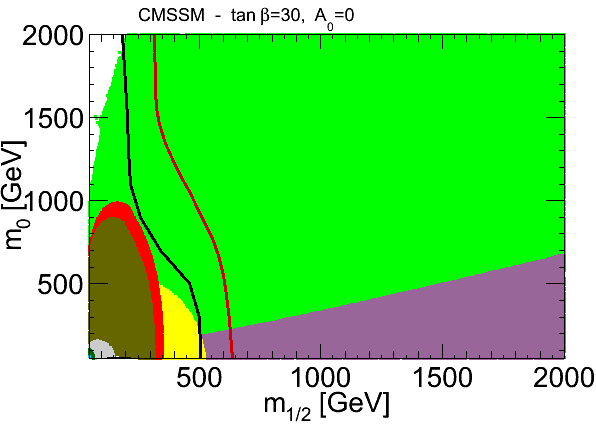}\quad\quad\raisebox{0.3cm}{\includegraphics[width=2.5cm]{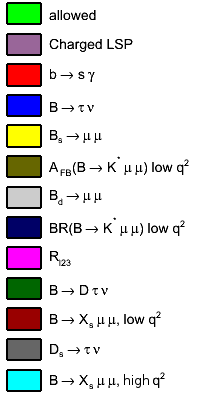}}
\caption{Constraints from flavour observables in CMSSM in the plane ($m_{1/2}, m_0$) for $\tan\beta=$ 30.}
  \label{Figure:Nazila2}
\end{center}
\end{figure*}
In an exemplary mode we show the strong restriction power of this data on the parameter space of the CMSSM as presented in Refs.~\cite{Akeroyd:2011kd,Mahmoudi:2012uk}. 
In Figure~\ref{Figure:Nazila1}, taken from Ref.~\cite{Mahmoudi:2012uk}, constraints from flavour observables on the CMSSM in the plane ($m_{1/2}, m_0$) for {a typical large $\tan\beta$ scenario with} $\tan\beta= 50$ and $A_0=0$, are shown, in the left with the 2011 results for BR($B_s\to \mu^+\mu^-$), and in the right with the {2012 Moriond} results. The colour code is as in Figure~\ref{Figure:Nazila2}. 
The black line corresponds to the CMS SUSY exclusion limit with 1.1 fb$^{-1}$ of data~\cite{Chatrchyan:2011zy} and the red line to the CMS SUSY exclusion limit with 4.4 fb$^{-1}$ of data~\cite{CMS-PAS-SUS-12-005} {at 7 TeV}. 
One notices that while with more integrated luminosity the direct limit is slightly shifted to higher masses, the constraining power of the new BR($B_s\to \mu^+\mu^-$) limit has {impressively} increased and overpassed the direct limit for high \mbox{values of $\tan\beta$.} 

Figure~\ref{Figure:Nazila2} shows that while the rare decay BR($B_s\to \mu^+\mu^-$) is very constraining in the large $\tan\beta$ region, it looses sensitivity when considering smaller values for $\tan\beta$. 
This conclusion does not change when considering more general MSSM scenarios with no universality assumption imposed. The sensitivity of the $B_s\to \mu^+\mu^-$ rate is significant in specific regions of the SUSY parameter space, mostly at large values of $\tan\beta$. As a result, as shown in~\cite{Arbey:2012ax}, the current LHCb measurement, and even foreseen future improvements in its accuracy, will leave a major fraction of the SUSY parameter space, compatible with the results of direct searches, unconstrained.
However, if a SUSY particle is discovered in direct searches at ATLAS and CMS, the precise value of BR($B_s\to \mu^+\mu^-$) would be very important for consistency checks, and could be used to severely constrain the parameters and help discriminating between different hypotheses.

\section{Latest news from the $\boldsymbol{B}$ factories}
\subsection{News on inclusive penguins? }
The inclusive decay $\bar B \to X_s \gamma$ is a good example to confirm  the simple CKM theory of flavour mixing in the SM, not
shown in the CKM  unitarity fit.
While non-perturbative  corrections to this decay mode are subleading and recently estimated to be well below $10\%$~\cite{Benzke:2010js}, 
perturbative QCD corrections are the most important corrections.
Within a global effort, a perturbative QCD calculation to the next-to-next-to-leading-logarithmic 
order level (NNLL) has been performed and has led to the first NNLL prediction of the $\bar B \to X_s  \gamma$ branching fraction~\cite{Misiak:2006zs} with a photon cut at $E_\gamma = 1.6$ GeV (including the error due to nonperturbative corrections):
\begin{equation}\label{final1}
\mathrm{BR}(\bar B \to X_s \gamma)_{\rm NNLL} =  (3.15 \pm 0.23) \times 10^{-4}.
\end{equation}
Using updated input parameters from PDG in particular for the quark masses and the CKM elements, the central value is shifted to $3.08 \times 10^{-4}$.
The combined experimental data by HFAG leads to~\cite{Amhis:2012bh} 
\begin{equation}
 {\mathrm{BR}}(\bar B \rightarrow X_s \gamma) = (3.43 \pm 0.21 \pm 0.07) \times 10^{-4}, 
\end{equation}
where the first error is combined statistical and systematic, and the second
is due to the extrapolation in the photon energy.
Thus, the SM prediction and the experimental average are consistent at the $1.2 \sigma$ level.
As a consequence, the $\bar B \to X_s \gamma$ has very  restrictive power on the parameter space of NP models. 
Recently, the first practically complete NLL calculation of this decay in the MSSM has been finalised~\cite{Greub:2011ji,Greub:2011qt}. 
\begin{figure}[t!]
\includegraphics[width=22pc]{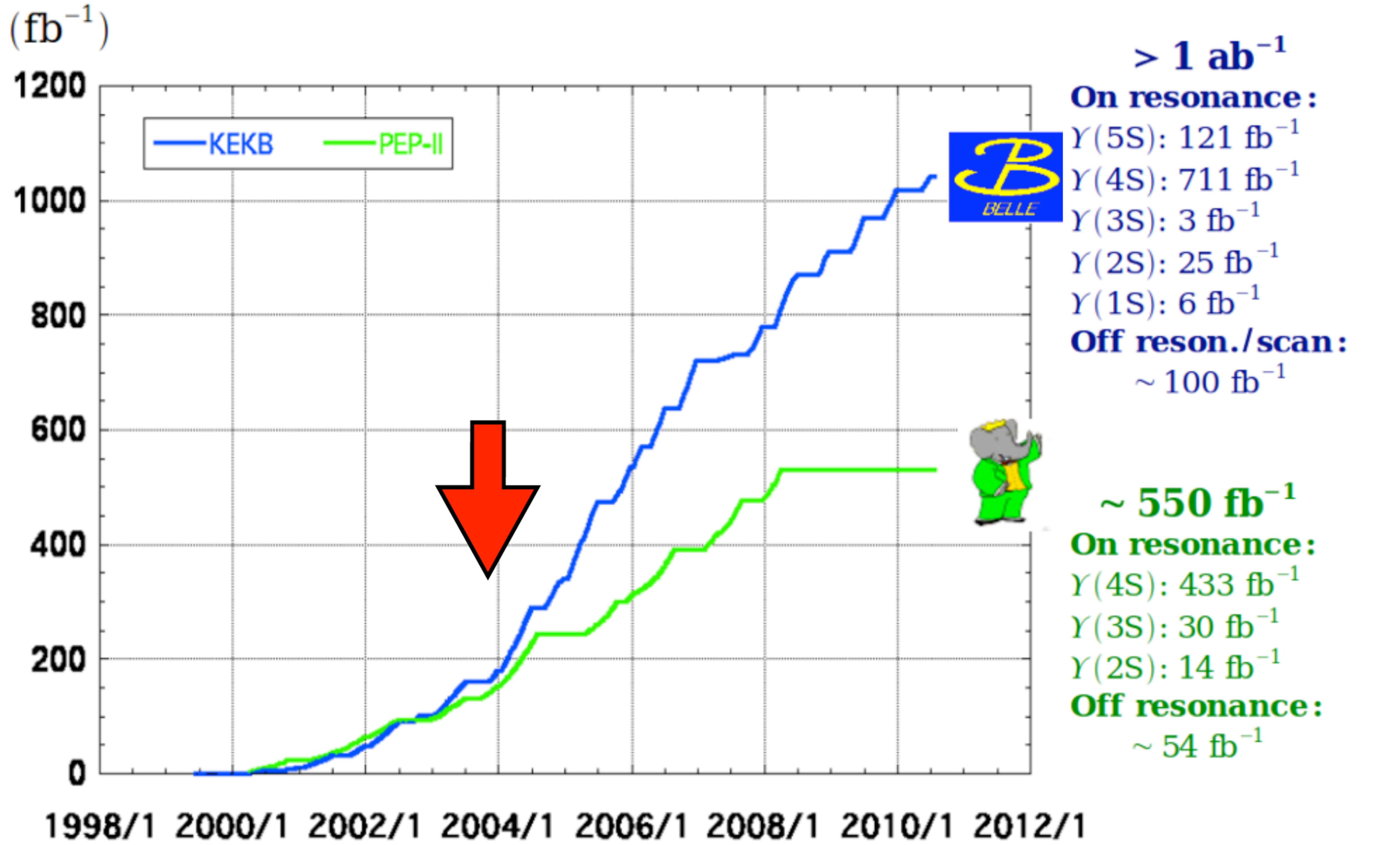}
\caption{\label{Integrated}Integrated luminosity of the $B$ factories.}
\end{figure}
\begin{figure*}[!t]
\begin{center}
\includegraphics[width=.4\textwidth]{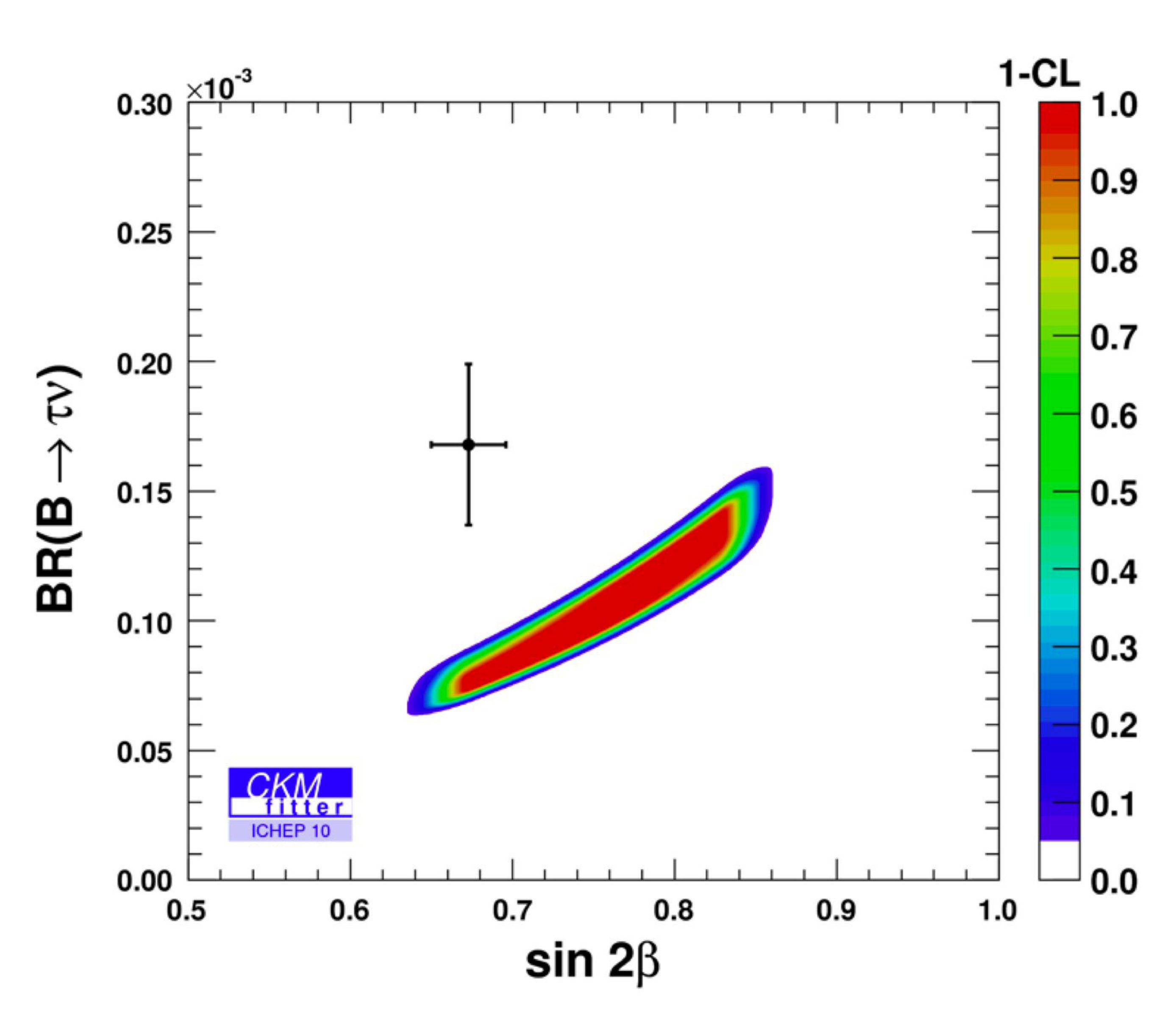}\quad\includegraphics[width=.4\textwidth]{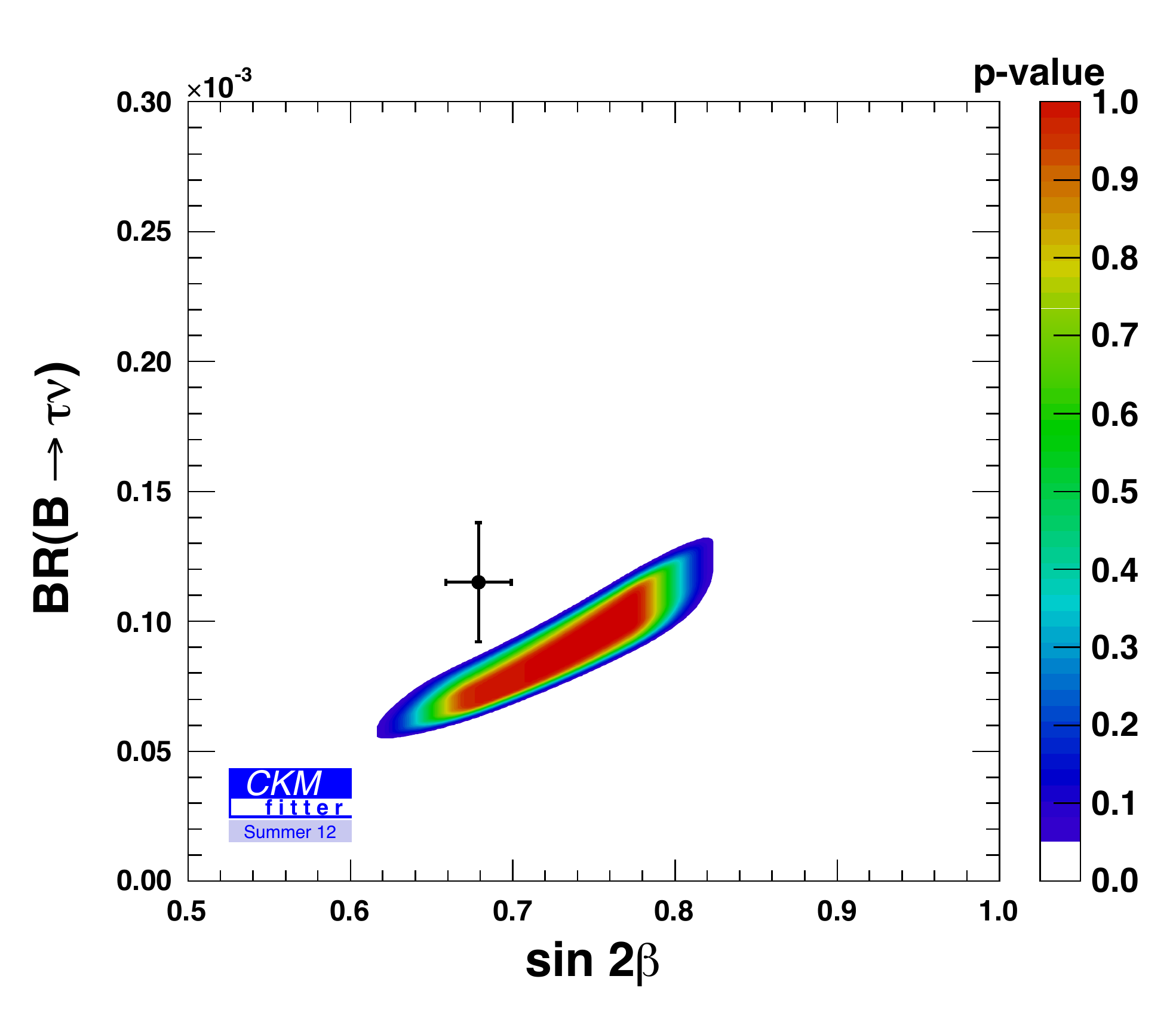}
\caption{Correlation of  ${\mathrm{BR} }(B \rightarrow \tau\nu)$ and $\sin\beta$ based on Pre-ICHEP12 data (left) and on ICHEP12 data (right); the cross corresponds to the 
experimental  values with $1 \sigma$ uncertainties~\cite{CKMfitter}.\label{fig:CKMfitter12}}
\end{center}
\end{figure*}
\begin{figure*}[!t]
\begin{center}
\includegraphics[width=.30\textwidth]{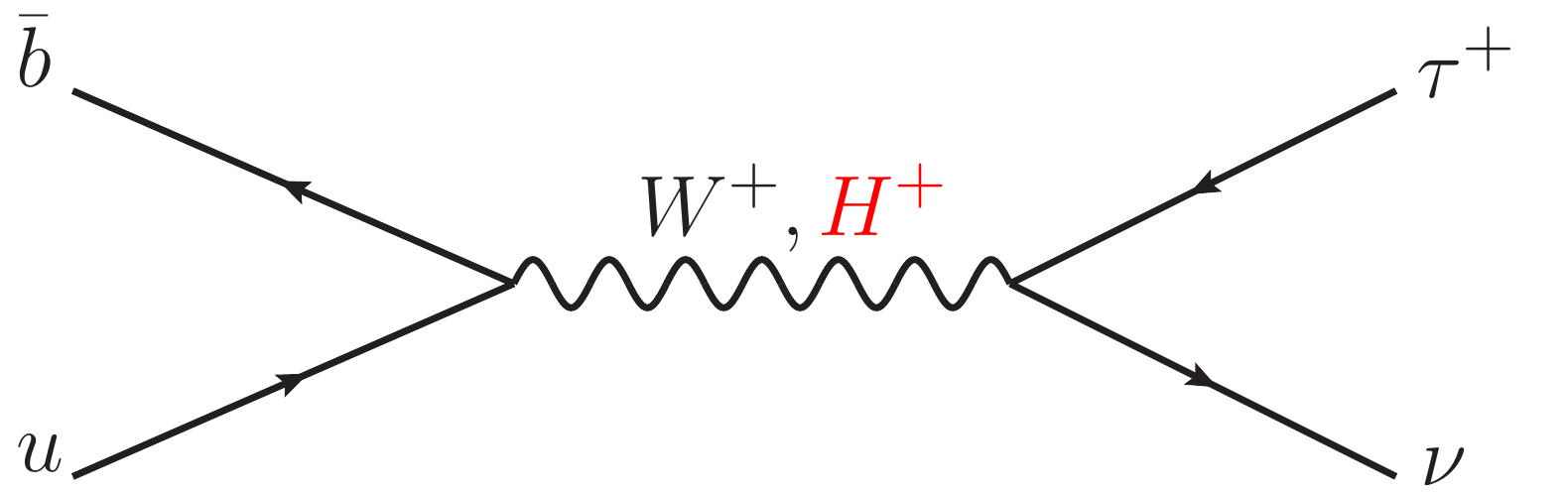}\quad\includegraphics[width=.30\textwidth]{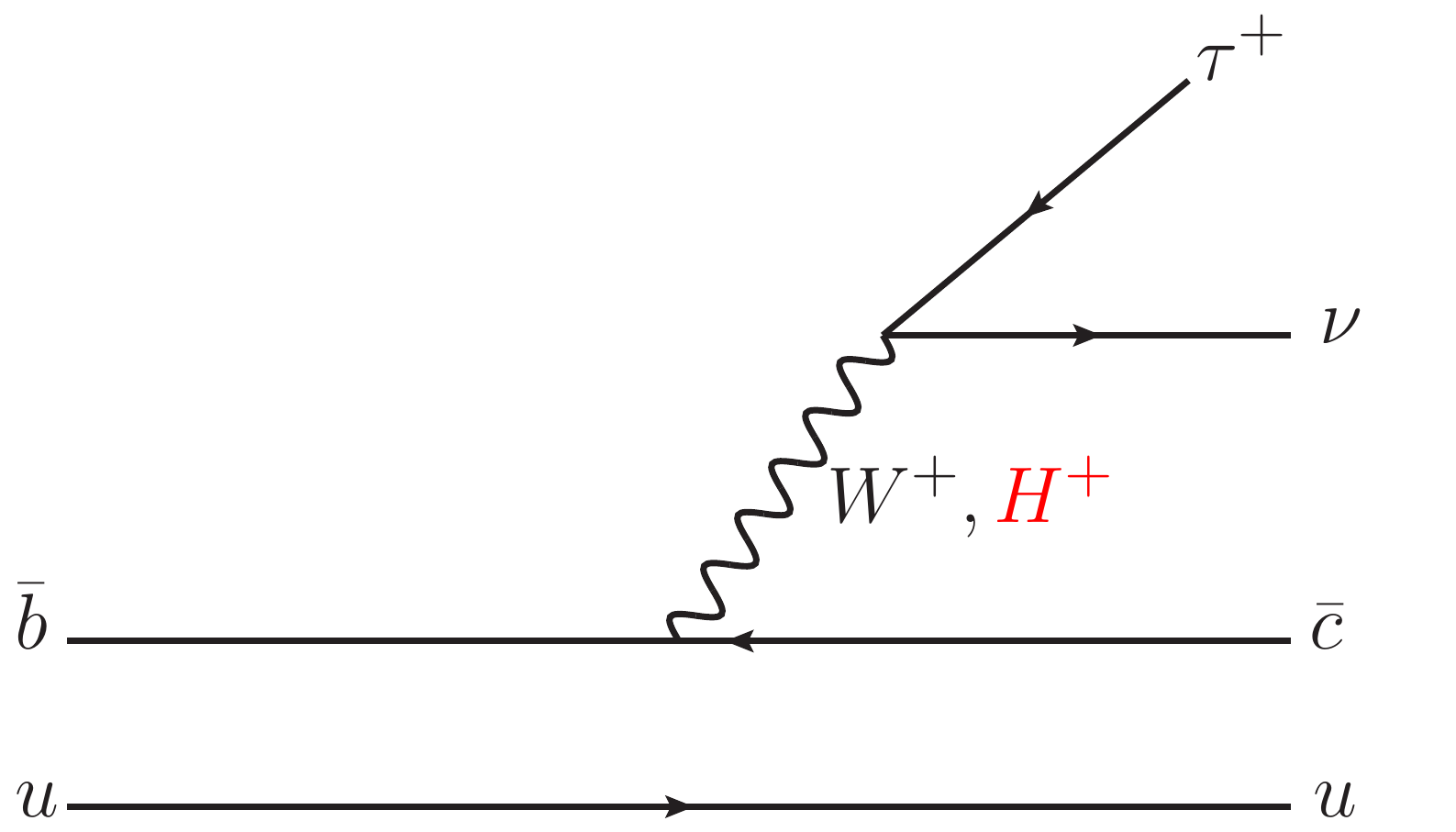}
\end{center}
\caption{Tree contributions to ${\mathrm{BR}}(B \rightarrow \tau \nu)$ (left) and to ${\mathrm{BR}}(B \rightarrow D \tau\nu)$ (right).}
\label{fig:feynman45}
\end{figure*}

The inclusive semi-leptonic decay $B \to X_s \ell^+ \ell^-$ could in principle play a similar role in the NP search. The NNLL QCD calculation 
has been finalised some time ago and even electromagnetic corrections have been calculated recently. The theoretical accuracy 
is  of order of $10\%$~\cite{Huber:2007vv}. However, unfortunately the {\it latest} measurements  of this inclusive decay mode of the $B$ factories 
stem from 2004 in case of BaBar based on $89 \times 10^6 B  \bar B$ events~\cite{Aubert:2004it}  and from 2005  in case of Belle based on 
$152 \times 10^6 B \bar B$ events~\cite{Iwasaki:2005sy}.
As the graph of the integrated luminosity {(Figure~\ref{Integrated})} shows these numbers of events correspond to less than $30\%$ of the data set available at the end 
of the $B$ factories. It would be highly desirable that new analyses are worked out which are based on the complete data sets.  
For further details on inclusive penguin decays we refer the reader to the recent mini-review on penguins~\cite{Hurth:2011jw}.

\subsection{New physics  in $\boldsymbol{B \to \tau \nu}$?}\label{Btaunu}
For some  time there has been a tension between the direct measurement and the indirect fit of the branching ratio ${\mathrm{BR}}(B \to \tau \nu)$
at the $2.8 \sigma$ level.  
Moreover, as was pointed out by the CKMfitter group~\cite{CKMfitter}, there has been a specific correlation 
between $\sin \beta$ and \mbox{${\mathrm{BR}}(B \rightarrow \tau\nu)$}  which is also a bit at odds, see Figure~\ref{fig:CKMfitter12}. 
Obviously the measured value of $\sin\beta$ has been too low, while the one of ${\mathrm{BR}}(B \to \tau \nu)$ has been  too large.
Interestingly, this tension could have been solved by a negative NP mixing phase in the $B_d$ system, $\Phi_d^\Delta  <  0$.

In principle, one could think  that this tension could also be solved by a NP contribution to ${\mathrm{BR}}(B \rightarrow \tau\nu)$
induced by a charged Higgs in the popular Two-Higgs Doublet model {(2HDM)} of Type II, see left diagram in Figure~\ref{fig:feynman45}. 
In the later model the SM branching ratio gets modified
in the following way:
\begin{equation}
{\mathrm{BR}}(B \to \tau \nu) = {\mathrm{BR}}_{SM} \times  \left( 1 - \frac{m_B^2}{M_{H^+}^2}\, \tan^2\beta \right)^2.
\end{equation}
But for the allowed values of the ratio of  the quantity $\tan\beta$ and the charged Higgs mass $M_{H^+}$  due to constraints by
other flavour data one only gets a reduction compared to the SM branching ratio.

\begin{figure}[htb]
\begin{center}
\includegraphics[width=.50\textwidth]{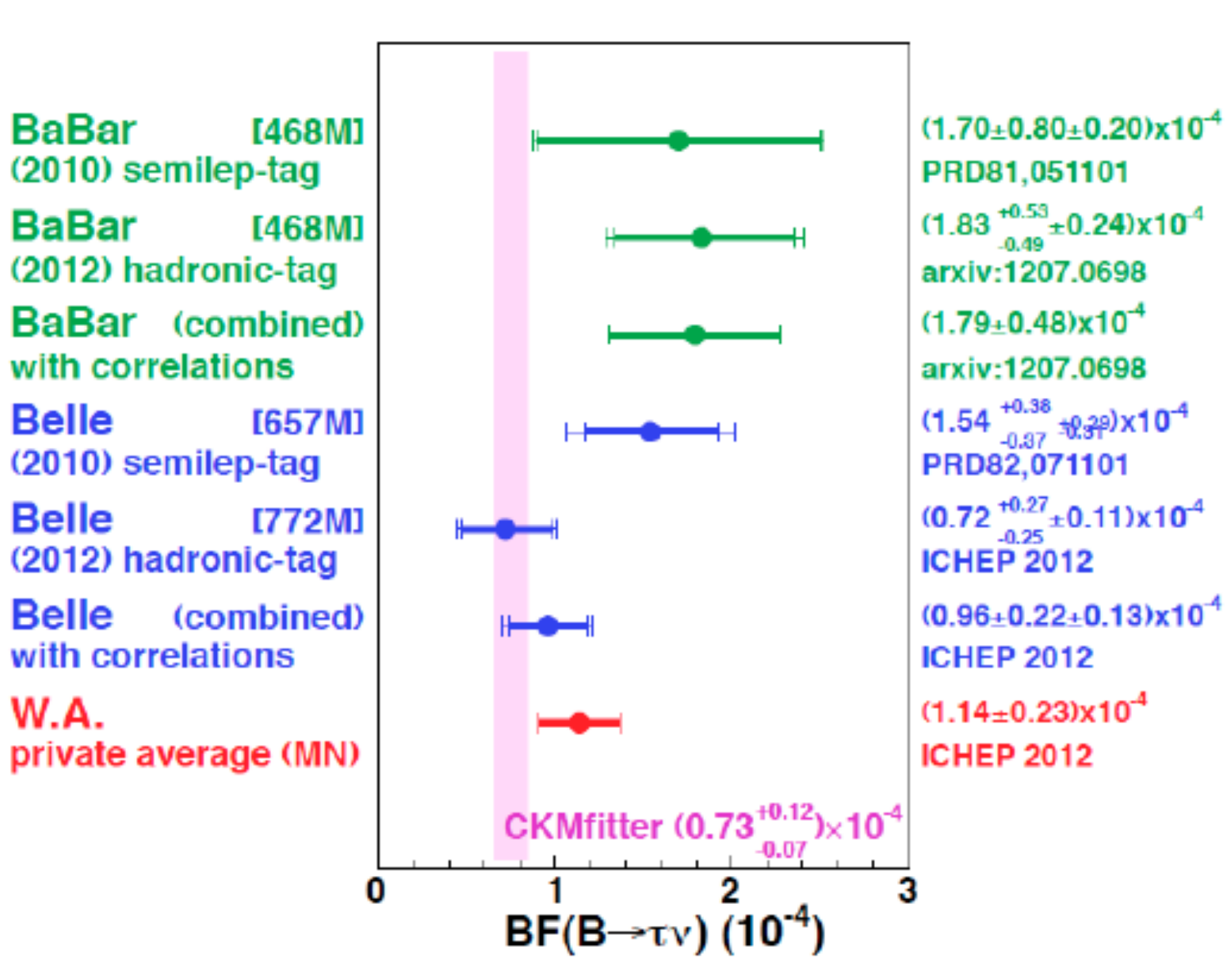}
\end{center}
\caption{ICHEP12 data: various measurements of  ${\mathrm{BR}}(B \rightarrow \tau \nu)$ with the new world average, courtesy of M. Nakao.}
\label{fig:CKMfitter13}
\end{figure}
However, Belle has recently  presented a new measurement with new data and an improved  analysis method including also a reanalysis of the old data
which shows a significant lower value in good agreement with the global fit,  while the new Babar measurement  confirms  the old high value. The various measurements 
are shown in Figure~\ref{fig:CKMfitter13}. 
As a result the indirect fit prediction for   ${\mathrm{BR}}(B \rightarrow \tau \nu)$  and direct measurements presently deviate by $1.6\sigma$ only, see Figure~\ref{fig:CKMfitter12}.

More recently a similar tension has showed up in $B \to D \tau \nu$ and $B \to D^* \tau \nu$.  Based on its  full  data sample,  Babar has reported  an improved 
measurements of the specific ratios~\cite{Lees:2012xj}:
\begin{eqnarray}
{\cal R}_{\tau/\ell} &=& {\mathrm{BR}}(B \to D \tau \nu)/{\mathrm{BR}}(B \to D \ell \nu)\\
 &=&   0.440 \pm 0.058 \pm 0.018   \nonumber\\
{\cal R}^*_{\tau/\ell} &=&  {\mathrm{BR}}(B \to D^* \tau \nu)/{\mathrm{BR}}(B \to D^* \ell \nu)\\
 &=&  0.332 \pm 0.024 \pm 0.018.  \nonumber
\end{eqnarray}
They exceed the SM expectations by $2.0\sigma$ and $2.7\sigma$, respectively~\cite{Fajfer:2012vx,Becirevic:2012jf,Bailey:2012jg}.

These ratios are rather sensitive to new physics contributions because the hadronic form factors tend to cancel. For example,  they  are   
sensitive to the  charged Higgs, see right diagram in Figure~\ref{fig:feynman45}. But again the 2HDM-II  does not offer a consistent explanation of the two ratios;
for the allowed values of $\tan\beta/M_{H^+}$, one finds an explanation for ${\cal R}$ but not for ${\cal R}^*$.  As shown in Ref.~\cite{Crivellin:2012ye}, 
a consistent explanation of both ratios is possible in the 2HDM of type III.  Interestingly,  the authors of Ref.~\cite{Fajfer:2012jt} argue that MFV (see next section) is disfavoured as explanation of this anomaly and spot various models with general flavour structures for it. Since the current result still suffers from large systematic uncertainty due to the background, the updated BaBar results and confirmation from Belle are awaited to clarify the situation.

\section{MFV benchmark}
At this stage of the NP search using rare $B$ and kaon decays, it makes sense  to analyse the impact of the measurements within the  framework 
of minimal flavour violation (MFV).  
The hypothesis of MFV~\cite{Chivukula:1987py,Hall:1990ac,D'Ambrosio:2002ex,Hurth:2008jc}, 
is a formal model-independent  solution to the 
NP flavour problem. It assumes that the flavour and the CP 
symmetries are  broken as in the SM.  Thus, it 
requires that all flavour- and CP-violating interactions be 
linked to the known structure of Yukawa couplings.     
A renormalisation-group invariant definition of MFV 
based on a symmetry principle 
is given in Ref.~\cite{D'Ambrosio:2002ex}; 
this  is mandatory for  a consistent  
effective field theoretical analysis
of NP  effects (for a recent mini-review see Ref.~\cite{Isidori:2012ts}). 

\begin{figure*}[t!]
\begin{center}
\includegraphics[width=5.6cm]{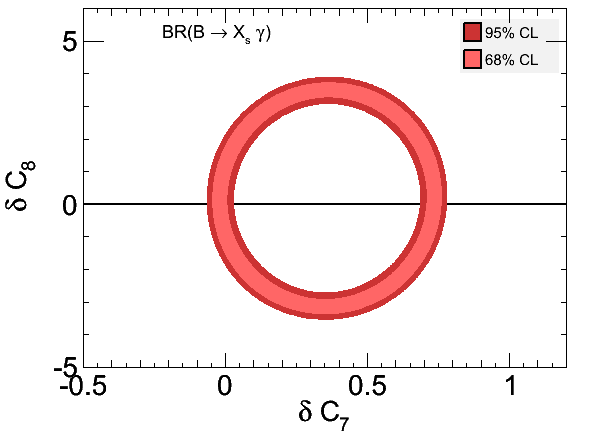}\quad\quad\includegraphics[width=5.5cm]{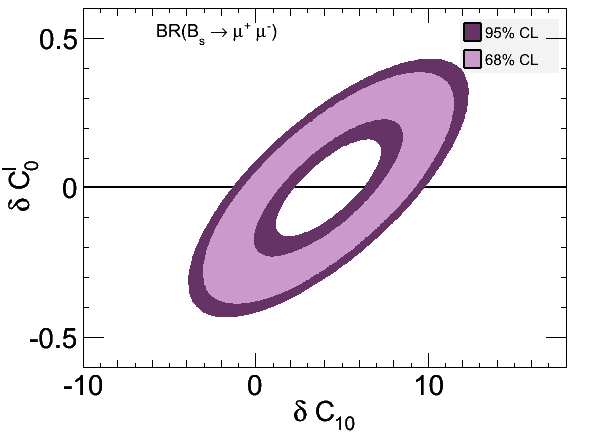}
\caption{ 68\% and 95\% C.L. bounds on $\delta C_7$ and $\delta C_8$ induced by the inclusive decay $\bar B\to X_s \gamma$ (left) and on $\delta C_{10}$ and $\delta C_0^\ell$ induced by the decay $B_s\to \mu^+\mu^-$ (right).}
\label{MFV-ind}
\end{center}
\end{figure*}

The MFV hypothesis represents an  important benchmark in the sense that any measurement  which 
is inconsistent with the general constraints and relations \mbox{induced} by the MFV 
hypothesis  unambiguously indicates the existence of new flavour structures. 
Moreover, compared with a general model-independent analysis as presented in Ref.~\cite{Altmannshofer:2012az,DescotesGenon:2011yn,Beaujean:2012uj}, the number of free parameters is heavily reduced 
due to the additional MFV relations. 
Indeed there are two strict predictions in this general 
class of models  which have to be tested.  First, the MFV hypothesis implies 
the  usual CKM relations between $b \to s$, $b \to d$, 
and  $s \to d$ transitions. For example, this relation allows 
for upper bounds  on NP  effects in 
${\rm BR}(\bar B \to X_d\gamma)$, and ${\rm BR}(\bar B \to X_s \nu\bar \nu)$ using experimental data 
or bounds from ${\rm BR}(\bar B  \to X_s\gamma)$, and 
${\rm BR}(K \to \pi^+ \nu\bar \nu)$, respectively. 
This emphasises the need for 
high-precision measurements of $b \to  s/d$ , but also of 
$s \to  d$ transitions such as  the rare kaon decay 
$K \to \pi \nu\bar\nu$. 
The second prediction is  that  the CKM phase is the only   
source of CP violation. This implies that any phase 
measurement as in 
$B \to \phi K_s$  is not sensitive 
to  new physics. 
This is an  additional assumption because the breakings of the flavour group and 
the discrete CP symmetry are in principle not connected at all. 
For example there is also a
renormalisation-group invariant extension of the MFV concept allowing for flavour-blind phases 
as was shown in Ref.~\cite{Hurth:2003dk}; however these lead to nontrivial CP  effects,  
which get strongly constrained by flavour-diagonal observables 
such as electric dipole moments~\cite{Hurth:2003dk}. 
So within the model-independent effective field theory approach of MFV we keep
the minimality condition regarding CP. 
But in specific models like MSSM the \mbox{discussion} of \mbox{additional} CP phases within the MFV
framework makes sense and can also allow  for  a natural solution of the well-known 
supersymmetric CP
problem, see for example Refs.~\cite{Mercolli:2009ns,Paradisi:2009ey}.

The application of the MFV hypothesis to the MSSM offers two attractive features. Most interestingly, the MFV hypothesis can serve as a substitute for R-parity in the MSSM~\cite{Nikolidakis:2007fc,Csaki:2011ge}. MFV  is sufficient to forbid a  too fast proton decay because when the MFV hypothesis is applied to R-parity violating terms, the spurion expansion leads to a suppression by neutrino masses and light-charged fermion masses, in this sense MFV within the MSSM  can be regarded as a natural theory for R-parity violation. 
Secondly, the MFV framework is renormalisation-group invariant by construction, however, it is not clear that the hierarchy  between the spurion terms is preserved when running down 
from the high scale to the low electroweak scale. Without this conservation of hierarchy, the MFV hypothesis would lose its practicability. However, as explicitly shown in 
Refs.~\cite{Paradisi:2008qh,Colangelo:2008qp}, a MFV-compatible change of the boundary conditions at the high scale has barely any influence on the low-scale spectrum.    

It is worth mentioning that the MFV hypothesis solves the NP flavour problem only formally.   One still has to find explicit dynamical structures to realise the MFV hypothesis like gauge-mediated supersymmetric theories. And of course  the MFV hypothesis is not a theory of flavour; it does not explain the hierarchical structure
of the CKM matrix and the large mass splittings of the SM fermions.

We stress that the MFV hypothesis is far from being verified. There is still room for sizeable new effects, 
and  new flavour structures beyond the Yukawa couplings are still compatible 
with the present data because the flavour sector has been tested only 
at the $10\%$  level  especially in the $b\to s$  transitions.

Based on the recent LHCb data a new analysis of rare decays within the MFV effective theory was presented~\cite{Hurth:2012jn}. {Here we update that analysis using the latest LHCb result for BR($B_s \to \mu^+\mu^-$) and the new HFAG world average for BR($B \to X_s \gamma$).}

Within the MFV effective Hamiltonian one singles out only five relevant  $b \to s$ operators  (and also $b \to d$ operators 
with obvious replacements):
\begin{eqnarray}
{\cal H}^{ b\to s}_{\rm eff} &=& -\frac{4 G_F}{\sqrt{2}}
[  V^*_{us} V_{ub} (C^c_1 P^u_1 + C^c_2 P^u_2)\\
&&  + V^*_{cs} V_{cb} (C^c_1 P^c_1 + C^c_2 P^c_2)]
\nonumber \\
&-& \frac{4 G_F}{\sqrt{2}}    \sum_{i=3}^{10} [(V^*_{us} V_{ub}
+ V^*_{cs} V_{cb}) C^c_i \;\nonumber\\
&& + \; V^*_{ts} V_{tb} C^t_i] P_i
+
V^*_{ts} V_{tb}
C^\ell_{0} P^\ell_{0}~+~{\rm h.c.}\nonumber
\label{eq:newHeff}
\end{eqnarray}
where the relevant operators are
\begin{equation}
\begin{array}{lll}
P_7  &=&   \dfrac{e}{16\pi^2} m_b (\bar{s}_L \sigma^{\mu \nu}     b_R)
F_{\mu \nu}~,\\[2mm]
P_8  &=&   \dfrac{g_s}{16\pi^2} m_b (\bar{s}_L \sigma^{\mu \nu} T^a b_R)
G_{\mu \nu}^a~,\\[2mm]
P_9  &=&   \dfrac{e^2}{16\pi^2} (\bar{s}_L \gamma_{\mu} b_L) \sum_\ell
 (\bar{\ell}\gamma^{\mu} \ell)~,\\[2mm]
P_{10} &=&  \dfrac{e^2}{16\pi^2} (\bar{s}_L \gamma_{\mu} b_L) \sum_\ell
                             (\bar{\ell} \gamma^{\mu} \gamma_5 \ell)~,\\[2mm]
P^\ell_{0} &=& \dfrac{e^2}{16\pi^2} (\bar s_L b_R) (\bar \ell_R \ell_L)~.
\end{array}
\end{equation}
The NP contributions to the corresponding Wilson coefficients can be parametrised as:
\begin{equation}
\delta C_i = C_i^{\rm MFV} - C_i^{\rm SM}\;.
\end{equation}
We scan over $\delta C_7$, $\delta C_8$, $\delta C_9$, $\delta C_{10}$ and $\delta C_0^\ell$ in order to obtain constraints on the Wilson coefficients based on the experimental results. Consecutively, for each point, the flavour observables are computed with the {\tt SuperIso} program~\cite{Mahmoudi:2007vz,Mahmoudi:2008tp}. The obtained values are compared to the experimental results by calculating the $\chi^2$ in the usual way
and the global fits are obtained by minimisation of the $\chi^2$. 

{The individual constraints from the new BR($\bar B\to X_{s} \gamma$) and BR($B_s\to \mu^+\mu^-$) results are displayed in Figure~\ref{MFV-ind}. As compared to the previous constraints in~\cite{Hurth:2012jn}, the region favoured by BR($\bar B\to X_{s} \gamma$) is only slightly shifted, and the constraints from the upper bound of BR($B_s\to \mu^+\mu^-$) weakened while the lower
bound now excludes the central region.}

\begin{table*}
\hspace*{-0.5cm}\scriptsize{\begin{tabular}{|l|l|l|l|}\hline
Observable & Experiment (post-LHCb) & Experiment (pre-LHCb) & SM prediction\\ \hline
BR($B_s \to \mu^+\mu^-$) & $(3.2~^{+1.4}_{-1.2}~^{+0.5}_{-0.3})\times 10^{-9}$ \cite{Aaij:2012ct} & $< 5.8 \times 10^{-8}$ \cite{Aaltonen:2007ad} &$(3.53 \pm 0.38) \times 10^{-9}$\\ \hline
$\langle dBR/dq^2(B \to K^* \mu^+ \mu^-) \rangle_{lq^2}$ & $(0.42 \pm 0.04 \pm 0.04)\times 10^{-7}$ \nocite{LHCb-CONF-2012-008}\citetext{LHCb, 2012a} &$(0.32 \pm 0.11 \pm 0.03)\times 10^{-7}$ \nocite{CDF-note-10047}\citetext{CDF, 2010} & $(0.47 \pm 0.27)\times 10^{-7}$\\ \hline
$\langle dBR/dq^2(B \to K^* \mu^+\mu^-) \rangle_{hq^2}$ & $(0.59 \pm 0.07 \pm 0.04)\times 10^{-7}$ \nocite{LHCb-CONF-2012-008}\citetext{LHCb, 2012a} & $(0.83 \pm 0.20 \pm 0.07)\times 10^{-7}$ \nocite{CDF-note-10047}\citetext{CDF, 2010} &$(0.71 \pm 0.18)\times 10^{-7}$\\ \hline
$\langle A_{FB}(B \to K^* \mu^+ \mu^-) \rangle_{lq^2}$ & $-0.18 \pm 0.06 \pm 0.02$ \nocite{LHCb-CONF-2012-008}\citetext{LHCb, 2012a} &$ 0.43 \pm 0.36 \pm 0.06$ \nocite{CDF-note-10047}\citetext{CDF, 2010}& $-0.06 \pm 0.05$\\ \hline
$\langle A_{FB}(B \to K^* \mu^+\mu^-) \rangle_{hq^2}$ & $0.49 \pm 0.06 \pm 0.05$ \nocite{LHCb-CONF-2012-008}\citetext{LHCb, 2012a} &$0.42 \pm 0.16 \pm 0.09$ \nocite{CDF-note-10047}\citetext{CDF, 2010}& $0.44 \pm 0.10$\\ \hline
$q_0^2 (A_{FB}(B \to K^* \mu^+ \mu^-))$ & $4.9 ^{+1.1}_{-1.3} \;\mathrm{ GeV}^2$ \nocite{LHCb-CONF-2012-008}\citetext{LHCb, 2012a} & -- & $4.26 \pm 0.34\; \mathrm{ GeV}^2$\\ \hline
$\langle F_{L}(B \to K^* \mu^+ \mu^-) \rangle_{lq^2}$ & $0.66 \pm 0.06 \pm 0.04$ \nocite{LHCb-CONF-2012-008}\citetext{LHCb, 2012a} & $0.50 \pm 0.30 \pm 0.03$ \nocite{CDF-note-10047}\citetext{CDF, 2010}& $0.72 \pm 0.13$\\ \hline
BR($B \to X_s \gamma$) & $(3.43 \pm 0.21 \pm 0.07)\times 10^{-4}$ \cite{Amhis:2012bh} &$ (3.43 \pm 0.21 \pm 0.07)\times 10^{-4}$ \cite{Amhis:2012bh} &$(3.08 \pm 0.24)\times 10^{-4}$\\ \hline
$\Delta_0(B \to K^* \gamma)$ & $(5.2 \pm 2.6)\times 10^{-2}$ \cite{Amhis:2012bh} &$(5.2 \pm 2.6)\times 10^{-2}$ \cite{Amhis:2012bh}& $(8.0 \pm 3.9)\times 10^{-2}$\\ \hline
BR($B \to X_d \gamma$) & $(1.41 \pm 0.57)\times 10^{-5}$ \nocite{delAmoSanchez:2010ae,Wang:2011sn}\citetext{del Amo Sanchez; Wang} &$(1.41 \pm 0.57)\times 10^{-5}$ \nocite{delAmoSanchez:2010ae,Wang:2011sn}\citetext{del Amo Sanchez; Wang} & $(1.49 \pm 0.30)\times 10^{-5}$\\ \hline
BR($B \to X_s \mu^+\mu^-)_{q^2\in[1,6] \rm{GeV}^2}$ & $(1.60 \pm 0.68)\times 10^{-6}$ \nocite{Aubert:2004it,Iwasaki:2005sy}\citetext{Aubert; Iwasaki} &$(1.60 \pm 0.68)\times 10^{-6}$ \nocite{Aubert:2004it,Iwasaki:2005sy}\citetext{Aubert; Iwasaki} & $(1.78 \pm 0.16)\times 10^{-6}$\\ \hline
BR($B \to X_s \mu^+\mu^-)_{q^2>14.4 \rm{GeV}^2}$ & $(4.18 \pm 1.35)\times 10^{-7}$ \nocite{Aubert:2004it,Iwasaki:2005sy}\citetext{Aubert; Iwasaki} &$(4.18 \pm 1.35)\times 10^{-7}$ \nocite{Aubert:2004it,Iwasaki:2005sy}\citetext{Aubert; Iwasaki} & $(2.19 \pm 0.44)\times 10^{-7}$\\ \hline
\end{tabular}}
\caption{Post- and pre-LHCb results for rare decays with the updated SM predictions \cite{Mahmoudi:2008tp}. $lq^2$ refers to $q^2\in[1,6]$ GeV$^2$ and $hq^2$ to $q^2\in[14.18,16]$ GeV$^2$. 
\label{tab:obs} }
\label{Inputobservables}
\end{table*}
\begin{figure*}[t!]
\begin{center}
\includegraphics[width=4.5cm]{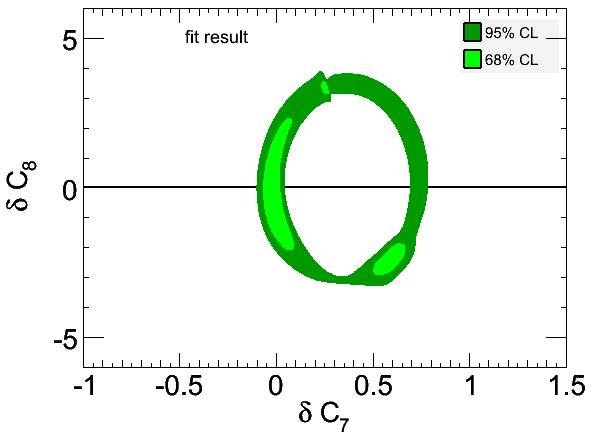}\includegraphics[width=4.5cm]{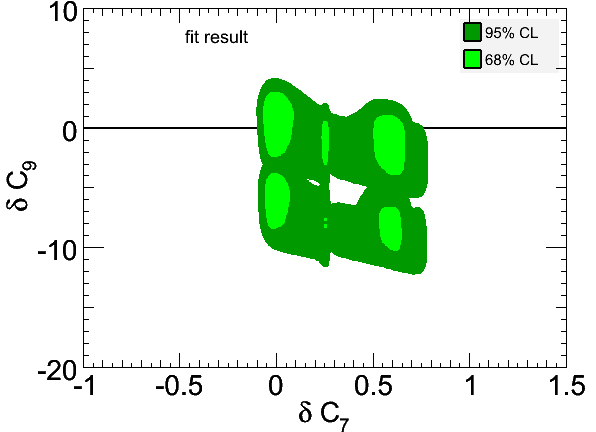}\includegraphics[width=4.5cm]{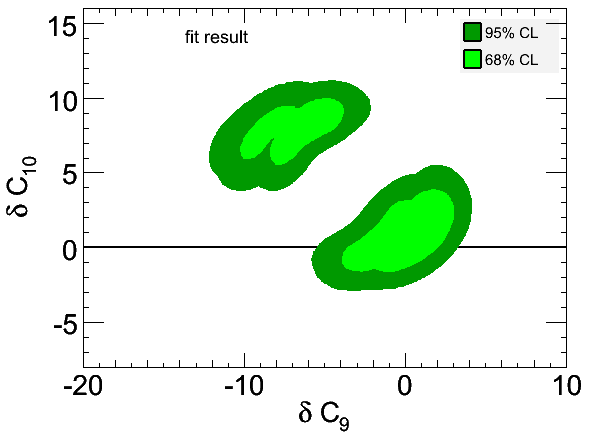}\includegraphics[width=4.5cm]{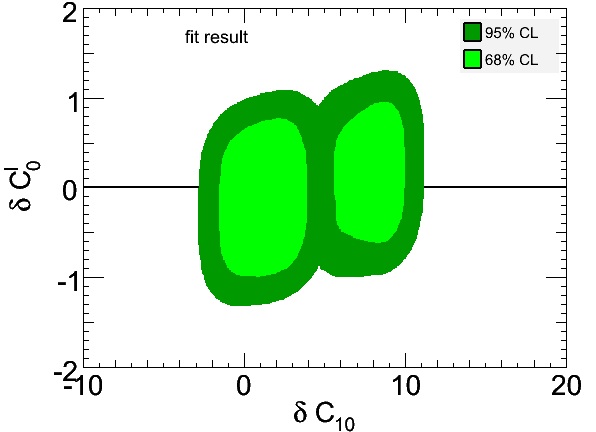}\\[0.3cm]
\includegraphics[width=4.5cm]{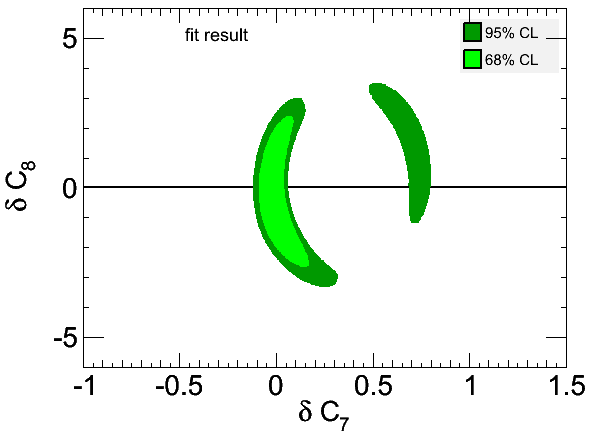}\includegraphics[width=4.5cm]{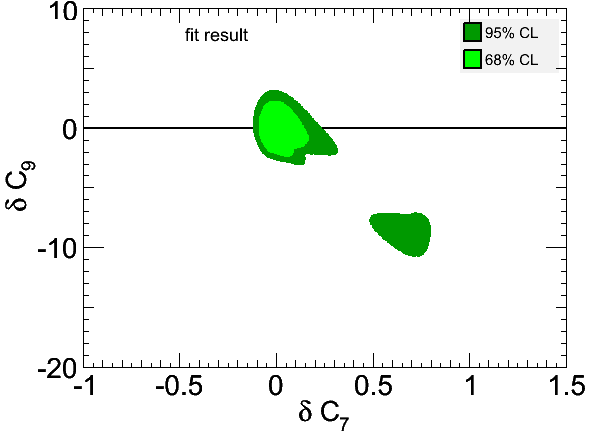}\includegraphics[width=4.5cm]{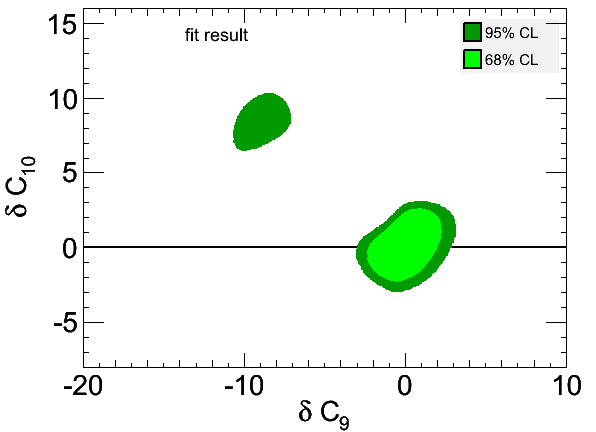}\includegraphics[width=4.5cm]{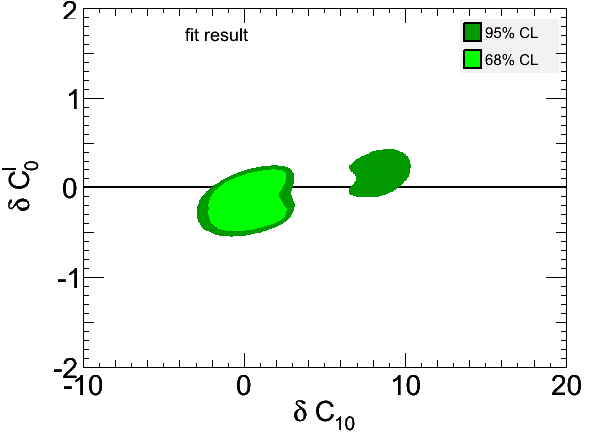}
\caption{Global MFV fit to the various NP coefficients $\delta C_i$ in the MFV effective theory {\it without} (upper panel) and  {\it with} experimental data of LHCb (lower panel).}
\label{MFVfit}
\end{center}
\end{figure*}

Two global MFV fits are given in Figure~\ref{MFVfit} to make the significance of the latest LHCb data manifest. In the first row, the experimental data before the start of the LHCb experiment are used (pre-LHCb), while the plots in the second row include the latest LHCb measurements (post-LHCb), as given in Table~\ref{tab:obs}. 
Here $C_8$ is mostly constrained by $\bar B\to X_{s,d} \gamma$, while $C_7$ is constrained by many other observables as well.
$C_9$ is highly affected by $b \to s \mu^+\mu^-$ (inclusive and exclusive). $C_{10}$ is in addition further constrained by $B_s\to \mu^+\mu^-$. The coefficient 
$C_0^\ell$ of the scalar operator is dominantly constrained by $B_s\to \mu^+\mu^-$. 
There are  always two allowed regions  at 95\% C.L. in the correlation plots within  the post-LHCb fit; one corresponds to SM-like MFV  coefficients and one to 
coefficients with flipped sign. The allowed region with the SM is more favoured. The various $\delta {C_i}$-correlation plots show the flipped-sign for $C_7$
is only possible if $C_9$ and $C_{10}$ receive large non-standard contributions which finally also change the sign of these coefficients.  
With the help of the results of the global fit, which restricts the NP contributions $\delta C_i$, we can now derive several interesting predictions 
for observables which are not yet well measured. This analysis also allows to spot the observables which still allow for relatively large deviations
from the SM (even in the MFV benchmark scenario). 
The following MFV predictions at the 95\% C.L. are of particular interest: 
\begin{eqnarray}
&1.0 \times 10^{-5}   \,<\,  {\rm BR}(\bar B \to X_d \gamma) \, < \, 4.0 \times 10^{-5} \,,\\ 
&{\rm BR}(B_d  \to \mu^+\mu^-)\, <\,  3.8  \times 10^{-10}.  
\end{eqnarray}
The present experimental results are \cite{Aaij:2012ct,delAmoSanchez:2010ae,Wang:2011sn}:
\begin{eqnarray}
&{\rm BR}(\bar B \to X_d \gamma)_{\rm Exp.} \, =  (1.41 \pm 0.57) \times 10^{-5} \,, \\
&{\rm BR}(B_d  \to \mu^+\mu^-)_{\rm Exp.} \, <\, 9.4  \times 10^{-10}.  
\end{eqnarray}
So the present $\bar B \to X_d \gamma$ measurement is already below the MFV bound and is nicely consistent with the correlation between the decays 
$\bar B \to X_s \gamma$ and $\bar B \to X_d \gamma$ predicted in the MFV scenario. 
In the case of the \mbox{leptonic} decay $B_d \to \mu^+\mu^-$, however, the MFV bound is stronger than the current experimental limit.
Moreover there are still sizeable deviations from the SM prediction possible within and also beyond the MFV bound  but
an enhancement by orders of magnitudes ({\it i.e.} due to large $\tan \beta$ effects) is already ruled out by the latest measurements.
Clearly, a measurement of $B_d \to \mu^+\mu^-$  beyond the MFV bound would signal the existence of new flavour structures beyond the Yukawa couplings.

\section{Outlook and future opportunities}
Many efforts have been deployed in the past in order to calculate as precisely as possible the low energy observables from flavour physics. This global effort led to a very satisfying situation now as we have access to several observables for which the theoretical predictions have reached high levels of accuracy. The reliability of the results from flavour physics (as compared to the other indirect searches such as in the dark matter sector where strong astrophysical and cosmological assumptions are needed) makes the flavour observables the premier actors in the search for indirect NP effects. Rare $B$ decays, and in particular $b \to s \gamma$ are the main assets here.
Also, the fact that multiple observables are available offers the opportunity for important cross checks.

In addition, any discovery at a high $p_T$ experiment must be consistent with the measurement from flavour experiments -- the contrary would indicate an inconsistency in the theory. The role of flavour physics is therefore very important in the LHC era. 

\begin{figure}[t!]
\begin{center}
\includegraphics[width=8.cm]{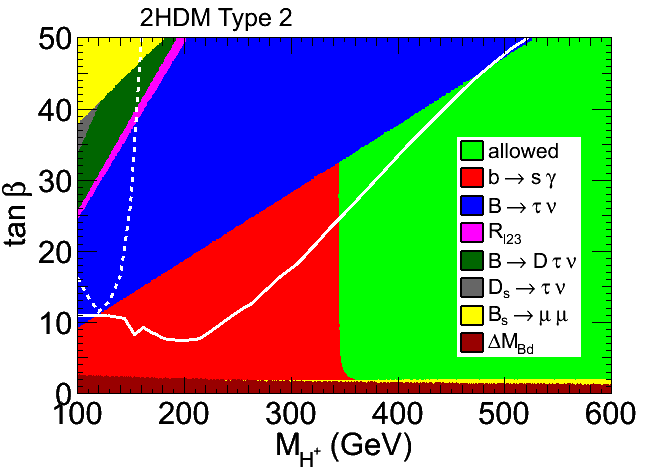}
\caption{Constraints from flavour observables in the 2HDM type II in the plane ($M_{H^+}$,$\tan\beta$). The green color indicates the parameter space still allowed by the flavour constraints. The constraints of the direct searches are indicated by the dashed or solid line (see text).  }
\label{2HDM}
\end{center}
\end{figure}

An example of the interplay between flavour constraints and LHC direct search results is displayed in Figure~\ref{2HDM} for the 2HDM type II, where BR$(b \to s\gamma)$ excludes the charged Higgs mass below 345 GeV for any value of $\tan\beta$. BR$(B \rightarrow \tau\nu)$ on the other hand constrains more strongly larger values of $\tan\beta$. These constraints can be compared to the latest limit from the direct searches of the charged Higgs boson by ATLAS \cite{Aad:2012tj} (dashed white line), where flavour constraints are clearly stronger, or with the CMS limit from direct $H/A\to\tau^+\tau^-$ searches \cite{Chatrchyan:2012vp} (solid white line);  {{here the CMS limit on $M_A$ has been transformed into a limit on $M_{H^+}$ assuming the tree-level MSSM mass relation $M_{H^+}^2 = M_A^2 + M_W^2$.}}. One notices the consistency and complementarity of the direct and indirect results.
Another concrete example is the understanding of the newly discovered Higgs-like particle properties where imposing consistency with the $b \to s \gamma$ and $B_s \to \mu^+ \mu^-$ results allows to discriminate between some of the underlying hypotheses.

We know that the stabilisation of the electroweak sector needs a nontrivial flavour structure which still has to be clearly identified. 
In spite of the fact that the first two years of high-statistics measurements of LHCb have not found any NP in FCNCs, still sizeable deviations from the MFV scenario are possible in various flavour observables. 
Thus, higher precision is needed to separate small deviations from the MFV benchmark. 

Also in other future scenarios for particle physics, flavour physics will be important. For example, in case no NP is discovered next to one scalar Higgs particle, the flavour precision experiments  may  show us the way to the NP energy scale. FCNCs provide indirect information about scales which are not accessible by the direct search.

There are  great experimental opportunities in flavour physics in the near future which will push the experimental precision to its limit.  
There are  $B$ physics programs at LHC at all three experiments  at CERN.
Especially LHCb will collect five times more data than the present data set.
The copious  production of all flavours of  $B$ mesons at the LHC, together with the unique
particle-identification capabilities of the LHCb detector, makes it possible to 
 investigate a wide range of decay channels that have  not
been accessible to previous  experiments. 
Most of them have been discussed in this report like  the CP-violating phase $\beta_s$, and searches of new physics effects via the rare decay modes $B \rightarrow K^* \mu\mu$
 and $B_s \rightarrow \mu\mu$, but 
also the measurement of the unitarity angle $\gamma$  and $B_s \rightarrow \phi\phi$.  An upgrade of the LHCb experiment  with a final integrated luminosity of 
$5$ fb$^{-1}$ to $50$ fb$^{-1}$ is planned and already approved~\cite{Merk:2011zz}.

There are also forthcoming experiments measuring rare $K$ decays such as
$K ^+ \rightarrow \pi^+  \nu \bar \nu$  and $K_L \rightarrow \pi^0 \nu \bar\nu$
(\citeauthor{CERNkaon,JPARCkaon}) which are extremely sensitive to possible new degrees of freedom and are largely unexplored.

In addition, two Super-$B$ factories, Belle II at KEK~\cite{Aushev:2010bq, Abe:2010sj}
and SuperB in Italy~\cite{O'Leary:2010af,Hitlin:2008gf,Bona:2007qt},  have been approved  and partially funded to accumulate two orders of magnitude larger data samples~\footnote{
The Italian government has recently decided that the latest cost estimate of the project is not compatible with the budget of the National Plan for Research.}.
The Super-$B$ factories are  actually Super-{\it Flavour} factories (SFF):  Besides precise $B$ measurements - 
for example,  the present  experimental error of  ${\mathrm{BR}}(B \rightarrow \tau\nu)$ discussed above will be reduced from $20\%$
down to $4\%$  improving the NP reach of this observable significantly  --
the SFF  allow for precise analyses of CP violation in charm and of lepton flavour-violating modes 
like $\tau \rightarrow \mu\gamma$ (see Ref.~\cite{Browder:2007gg}).
The results will be highly complementary to those on several 
important observables related to $B_s$ meson oscillations, 
kaon and muon decays that will be measured elsewhere.

Most important are the opportunities of a SFF  for lepton flavour physics. 
The sensitivity for  $\tau$ physics  is far superior to any other existing or proposed experiment, and the physics reach can be extended even further by the possibility to operate with polarised beams.
The study of the correlation of neutrino properties with flavour 
phenomena in the charged-lepton and in the quark sector, {\it e.g.} charged-lepton flavour violation, 
is also an important target. Pushing the present limits on $\mu \leftrightarrow e$
and $\mu \leftrightarrow \tau$ transitions might lead to important insight.
The combined information on $\mu$ and $\tau$ flavour violating decays that will be provided by the MEG experiment~(\citeauthor{MEG}) together with a SFF~\cite{Browder:2007gg}  may  shed light on the mechanism responsible for lepton flavour violation.  


\section*{Acknowledgement} TH 
thanks the CERN theory group for its hospitality during his regular visits to CERN where
part of this work was written.\\

\bibliographystyle{apsrmp4-1}
\bibliography{RMPsubmit2}

\end{document}